\documentclass[12pt]{report}
\usepackage[latin9]{luainputenc}
\usepackage{float}
\usepackage{booktabs}
\usepackage{amsmath}
\usepackage{graphicx}

\makeatletter

\providecommand{\tabularnewline}{\\}
\floatstyle{ruled}
\newfloat{algorithm}{tbp}{loa}[chapter]
\providecommand{\algorithmname}{Algorithm}
\floatname{algorithm}{\protect\algorithmname}

\newcommand{\reporttitle}{Comparison of Syntactic and Semantic Representations of Programs in Neural Embeddings}
\newcommand{\reportauthor}{Austin P. Wright}
\newcommand{\supervisor}{Herbert Wiklicky}
\newcommand{\degreetype}{Computing (Machine Learning)}
\newcommand{\HRule}{\rule{\linewidth}{0.5mm}} 

\date{August 2019}

\usepackage{tikz}
\usepackage{algorithmic}
\usepackage{adjustbox}

\makeatother

\usepackage{listings}

\begin{document}


\includegraphics[width = 4cm]{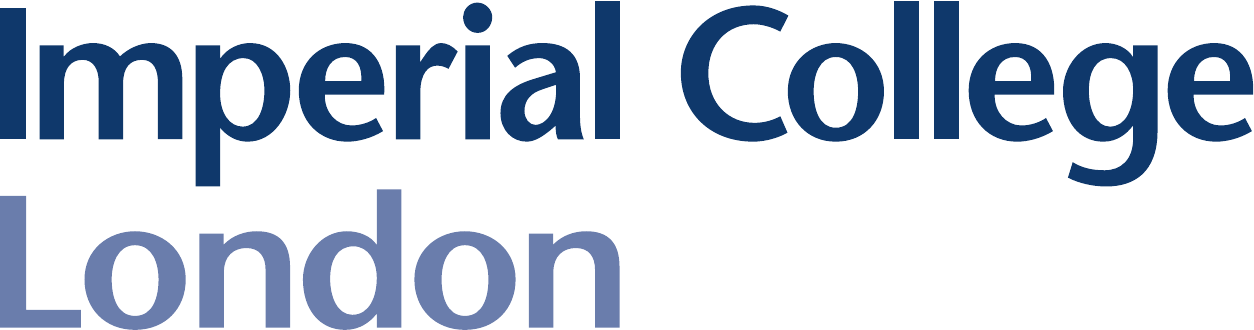}\\[0.5cm] 

\begin{center}


\textsc{\Large Imperial College London}\\[0.5cm] 
\textsc{\large Department of Computing}\\[0.5cm] 


\HRule \\[0.4cm]
{ \huge \bfseries \reporttitle}\\ 
\HRule \\[1.5cm]
 

\begin{minipage}{0.4\textwidth}
\begin{flushleft} \large
\emph{Author:}\\
\reportauthor 
\end{flushleft}
\end{minipage}
~
\begin{minipage}{0.4\textwidth}
\begin{flushright} \large
\emph{Supervisor:} \\
\supervisor 
\end{flushright}
\end{minipage}\\[4cm]

\vfill 
Submitted in partial fulfillment of the requirements for the MSc degree in
\degreetype~of Imperial College London\\[0.5cm]

\makeatletter
\@date 
\makeatother

\end{center} 

\clearpage\cleardoublepage{}
\begin{abstract}
Neural approaches to program synthesis and understanding have proliferated
widely in the last few years; at the same time graph based neural
networks have become a promising new tool. This work aims to be the
first empirical study comparing the effectiveness of natural language
models and static analysis graph based models in representing programs
in deep learning systems. It compares graph convolutional networks
using different graph representations in the task of program embedding.
It shows that the sparsity of control flow graphs and the implicit
aggregation of graph convolutional networks cause these models to
perform worse than naive models. Therefore it concludes that simply
augmenting purely linguistic or statistical models with formal information
does not perform well due to the nuanced nature of formal properties
introducing more noise than structure for graph convolutional networks.
\end{abstract}
\cleardoublepage{}

\cleardoublepage{}

\tableofcontents{}

\cleardoublepage{}

\chapter{Introduction}

Among the oldest goals in computer science has been the problem of
program synthesis, however it has also been among the most elusive.
Currently the most active area of interest in computer science is
deep learning, bringing groundbreaking new results in many problems
that were once very difficult. It is then natural to ask how we may
try to apply deep learning to the age old problem of program synthesis.

The current state of the art in program synthesis goes one of two
ways, either treating programs entirely like natural text in a single
sequence and thus using existing natural language processing techniques,
or eschewing deep learning altogether and using formal methods entirely
and treating programs as purely mathematical objects. One of the more
common and useful mathematical representations of programs is in various
kinds of static analysis graphs, such as a control flow graph. While
in the past these kinds of representations have not been compatible
with deep learning, recent developments in geometric deep learning
has opened the door to utilizing these graphs in neural networks.
The hope would be that a graph neural network would be able to have
all of the gains of a deep language system, and integrate the nuance
from mathematical analysis and combine this information for better
performance than either of the two other methods alone.

This work is an empirical study on the effectiveness of static analysis
graph based representations when compared to linguistic representations
of computer programs in deep learning systems. While the project proposal
was initially to study deep learning applications in program synthesis,
throughout this process and review it became clear that while there
have been certain conjectures on the effectiveness of graph based
representations, no study has been done to show that these kinds of
representations outperform other kinds of models. Towards answering
this question of model effectiveness in general this work considers
the specific case of program embedding, or encoding. This problem
is a useful formulation, as it is both useful in its own right within
a program synthesis framework like CEGIS (see section \ref{sec:Program-Synthesis}),
and can act as a proxy for other kinds of program understanding tasks
by forcing a model to learn to represent the whole program itself.
Therefore this work will analyze the comparative effectiveness of
equivalent models using either purely language based, or formal graph
augmented, program representations.

\chapter{Background}

Due to the breadth of topics discussed in this work, and the variance
in expertise of the relevant audience, no background in the relevant
fields of Deep Learning or Program Analysis are assumed beyond introductory
linear algebra and statistics. This background chapter aims to provide
relevant context and understanding of these fields required to follow
the later analysis; however it should not be taken as exhaustive or
supremely rigorous.

Section \ref{sec:Deep-Learning} on Deep Learning will introduce the
general machine learning problem formulation, and introduce the fundamental
pillars of how deep learning models work. It will then introduce the
specialized convolution and graph convolution structures that are
relevant to the models used in the work. Finally it will introduce
some basic natural language processing as well as static program analysis
techniques in order to understand how programs fit in to these deep
learning models. Generally more emphasis is given to the deep learning
portion of the background since bulk of the methods actually implemented
in the work involve deep learning, where as the program analysis techniques
used are rather simple. Overall this should provide enough background
for an undergraduate to get up to speed and understand the context
and content of the work outlined in the following chapters.

Section \ref{sec:Program-Synthesis}, while not directly relevant
to the specific methods used, includes a background and history of
Program Synthesis techniques. This is because this work began as an
analysis of these deep learning methods in program synthesis, and
the most relevant future works for these technologies is in that field.
Therefore this background while not technically relevant, aims to
inform the past and future context where the technologies analyzed
are most relevantly applied.

\section{\label{sec:Deep-Learning}Deep Learning}

\subsection{Machine Learning}

Since Deep Learning is just a special class of models in within the
general framework of machine learning we are first going to introduce
the fundamentals of machine learning. Machine Learning in effect is
the synthesis of statistical modeling which the reader is likely already
familiar, with a focus on computational tools and tractability. There
are general definitions for this class of algorithms that learn from
data such as the Probably Approximately Correct (PAC) Learning framework \cite{Valiant:1984:TL:1968.1972}.
In essence there are three main components in machine learning, the
\emph{data}, the \emph{model}, and the \emph{training algorithm. }The
fundamental machine learning task is for a given task, the \emph{model}
uses the \emph{training algorithm} on the \emph{data} in order to
improve its performance on said task. The most basic form of machine
learning would be the case of a linear regression, where given a set
of points (the data) find the line (linear model) that best matches
the trend in the data. 
\begin{figure}[ht]
\caption{Linear Regression}

\includegraphics[scale=0.1]{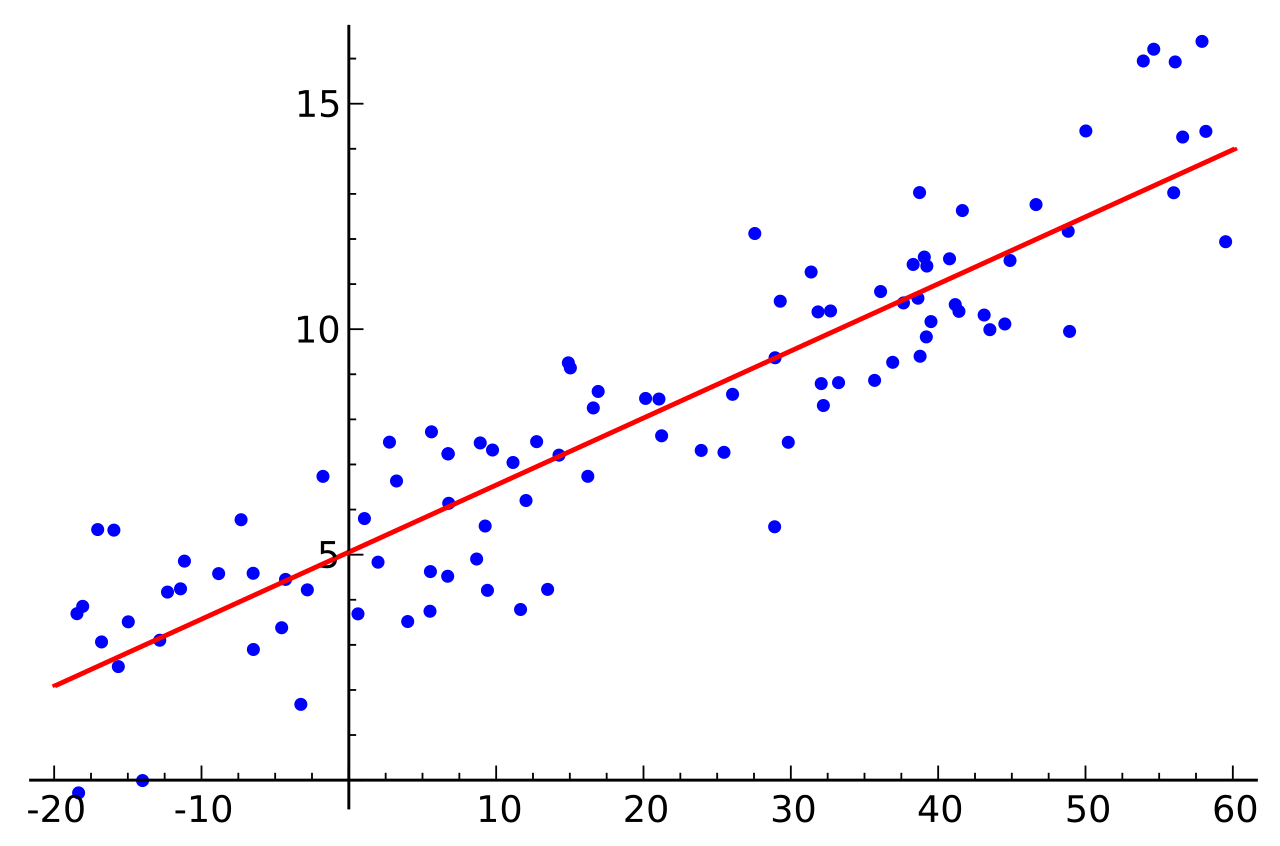}
\centering{}Source: https://en.wikipedia.org/wiki/File:Linear\_regression.svg
\end{figure}
This is done using the Ordinary Least Squares procedure (training
algorithm) in order to optimize a certain goal, in this case the square
error or total distance of the model line from the data. Often you
hear the term \emph{objective function,} this term simply represents
the metric by which the model is evaluated, which is essentially part
of the training algorithm (this is because much of the time the objective
function used to train a model might actually be different that whatever
final evaluation metric is used for a model).

There are a huge host of problems and types of data used for the machine
learning problem, since it is so general. However broadly these problems
can be broken into supervised learning and unsupervised learning.
In a supervised learning problem the data is divided into two parts,
the input and the label. In this situation normally a model is trained
to predict a label based on just an input. Supervised learning includes
everything from linear regression (as the model can be viewed as predicting
an output/label $y$ value from a given/input $x$ value), to machine
translation (predicting an output french translation from an input
english sentence). This is generally the most common form of machine
learning and performs the best as evaluation is easy with provided
ground truth in the form of labels. However often times this labeling
for the data is not possible or economical. Unsupervised Learning
is learning from data without labels. This includes clustering algorithms
that try and find structure in a dataset, or dimensionality reduction
like Principle Component Analysis that tries to represent a high dimensional
dataset with a small set of features. This framework is generally
less efficient in the information it can extract from a dataset since
there is no ground truth provided other than the data itself, however
this class can be very useful in the many cases where data labeling
is simply not an option.

The main goal of most machine learning models is to perform well not
just on a given dataset, but on new unseen data. This is called \emph{generalization}
and is a very hard problem. Different kinds of models have what is
called \emph{capacity} which essentially means the amount of flexibility
the model has in how fine tuned it is to small patterns in the data
(this generally correlates with the number of learnable parameters
in the model). 
\begin{figure}[ht]
\begin{centering}
\caption{Over and Under Fitting}
\par\end{centering}
\begin{centering}
 \begin{adjustbox}{center}\includegraphics[scale=0.4]{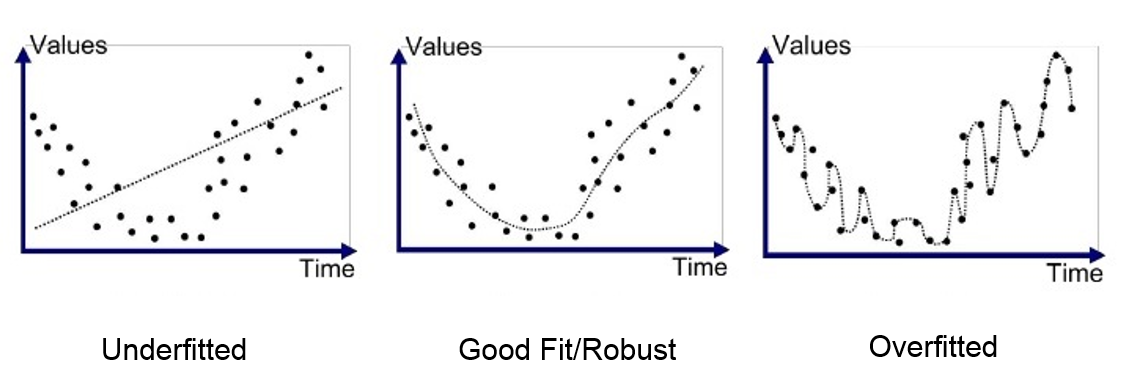}\end{adjustbox}
\par\end{centering}
\centering{}Source: https://medium.com/greyatom/what-is-underfitting-and-overfitting-in-machine-learning-and-how-to-deal-with-it-6803a989c76
\end{figure}
Very low capacity can lead to \emph{underfitting} where a model might
not be able to learn of of the most important patterns in the data.
However if a dataset has a lot of noise or the capacity of the model
is very high, a model could learn false patterns that are just results
of noise that will not generalize, this is called \emph{overfitting}.
Both of these effects hurt the generalization of a model, and so a
challenge in machine learning is finding the Goldilocks sweet spot
for a given problem between underfitting with too low of a capacity
and overfitting with too high. Controlling the model capacity is something
that is normally done through methods like \emph{regularization} which
penalize too high of a model capacity, or other methods that might
depend on the specific model at hand. We also attempt to measure this
generalization phenomena by a technique called \emph{cross-validation}
which will use a subset of the actual dataset to train, and then use
the held out portion to test the model on. This information can act
as a proxy for true generalization and is often used to tune \emph{hyperparameters}
or the parameters that are determined before training (things like
the size of the model or attributes of the training algorithm).

Deep Learning fits within this framework as simply a class of models
called Artificial Neural Networks. Later sections will go into more
depth on how these specific models work, but they still have the same
challenges as any other model might for machine learning problems.
They have fallen into favor as they have been shown to perform well
for a variety of very hard tasks (most importantly for this work text
and language processing), however it is important to know that deep
learning models are not fundamentally different than a linear regression
in the overall class of problems they are tryin to solve and the general
kinds of challenges that may arise.

\subsection{Multilayer Perceptron}

The most basic kind of deep learning model is the \emph{Deep Feedforward
Network} also called a \emph{Multilayer Perceptron}. However before
understanding what a multilayer perceptron is, perhaps we should first
cover what a single perceptron is. The perceptron is a computational
unit that takes inspiration from the biology of a neuron, in that
it has a set of inputs that are combined and then has what is called
an \emph{activation} which produces a single output. This is where
the similarities with biology and the brain for the most part end,
despite all of the hype around deep neural networks. 
\begin{figure}[ht]
\begin{centering}
\caption{Perceptron Unit}
\par\end{centering}
\begin{centering}
\includegraphics[scale=0.4]{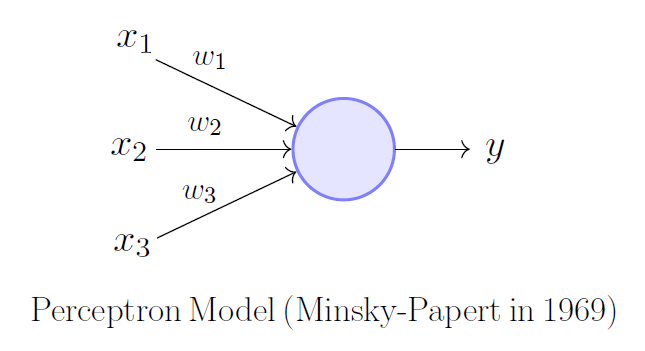}
\par\end{centering}
\centering{}Source: https://towardsdatascience.com/perceptron-the-artificial-neuron-4d8c70d5cc8d
\end{figure}
 Perceptrons actually date back to the earliest days of artificial
intelligence research in the sixties\cite{Minsky:1988:PEE:50066}.
Specifically for a given input vector $\vec{x}$ a perceptron will
take the dot product of $\vec{x}$ with a weight vector $\vec{w}$,
which is of course equivalent to taking a weighted sum of each individual
input. Then this output is passed into what is called an activation
function $\sigma$. The output thus of a perception is $y=\sigma(\vec{w}\cdot\vec{x})$.
In the early days this activation function was a simple threshold,
yielding $1$ if the input was large enough and $0$ otherwise. Later
this was adapted to a continuous version called the sigmoid function,
defined as:
\[
\sigma(x)=\frac{1}{1+e^{-x}}
\]

However today many different kinds of activation functions are used
including the hyperbolic tangent and the rectified linear unit (ReLU).
However regardless of activation function, a single perceptron is
not a particularly powerful model, it essentially is just a linear
model with an added function but cannot learn nonlinear features.
The key step in making perceptrons useful is stacking them together,
this puts the deep into deep learning and the multilayer in multilayer
perceptron. By chaining the outputs of perceptrons as the input to
another layer of perceptrons, this forms a ``hidden layer'' as the
inputs are processed to some intermediate value before finally being
processed into a final output ($y=\sigma_{2}(\vec{w_{2}}\cdot\sigma_{1}(\vec{w_{1}}\cdot\vec{x})$).
\begin{figure}[ht]
\begin{centering}
\caption{Multilayer Perceptron}
\par\end{centering}
\begin{centering}
\includegraphics[scale=0.3]{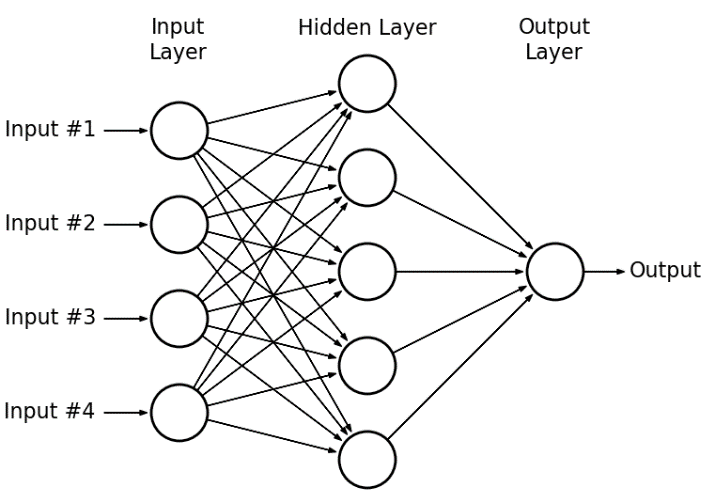}
\par\end{centering}
\begin{centering}
Source: https://www.researchgate.net/figure/A-hypothetical-example-of-Multilayer-Perceptron-Network\_fig4\_303875065
\par\end{centering}
\centering{}
\end{figure}
This architecture with just a single hidden layer actually forms on
its own a universal function approximator, where if there are a sufficient
number of neurons (perceptrons within a network are often called neurons)
in the hidden layer with a nonlinear activation function and a simple
linear activation for the final output layer, any function can be
modeled by this deep network (by learning the correct set of weights).
However the model is not limited to just one hidden layer, and empirically
it has been shown that generally adding more layers and making the
model ``deeper'' is more efficient than adding more neurons to a
given layer and making the model ``wider''.

\subsection{Training}

However a model with great ability to model functions in theory does
not matter much if there is no way to train it. That is the situation
deep feedforward networks were in for a long time until \emph{back-propagation}
was developed as a way to train these networks via \emph{gradient
descent}\cite{Rumelhart:1988:LRB:65669.104451}. Unlike linear
regression where parameters can be solved exactly in closed form,
the nonlinearities generally make the problem of finding the correct
weights in deep networks \emph{non-convex.} What that means is that
finding the correct weights is a much harder problem and involved
incrementally improving the model as opposed to calculating the solution
all at once. This is done by a method called \emph{gradient descent}
(descent because we are normally trying to minimize some kind of error
or loss). For a dataset we can calculate the loss of the model as
a function of the weights of the model, $L(\vec{\theta})$. Then in
order to change the weights and improve the loss the gradient (derivative)
is calculated $\nabla L(\vec{\theta})$ to find out a first order
approximation of how much the loss changes by changing each weight.
Then each weight is changed by a small amount in the direction that
would improve the loss. Then the process is repeated many times until
an optimum value of the weights is found. This process does not necessarily
guarantee a global optimum, however empirically if the process is
repeated enough times with different random starting values of the
weights it has been found that for the most part these local optima
are good. In practice this gradient is only calculated over a small
subset of the data (Stochastic Gradient Descent) and the size of the
update changes as the training process continues (ADAM), in addition
to a number of other small optimizations, however the core concept
remains the same.

This technique then comes down to calculating the gradient of the
loss in terms of each parameter. The process by which those values
are calculated is called back-propagation (although it is really just
application of the chain rule). The chain rule states that we can
take the derivative of composed functions $\left(\frac{d}{dx}z(y(x))=\left(\frac{d}{dy}z(y)\right)\cdot\left(\frac{d}{dx}y(x)\right)\right)$.
Since the output of each layer is just a function of the previous
layers we can use the chain rule to expand the $\nabla L(\vec{\theta})$
term in terms of the output of previous layers, and propagate that
information backwards through the network without having to recompute
partial derivatives. This work will not go into the specifics of how
the mathematics works, but suffice to say that this idea, while simple,
paved the way for training deep networks by gradient descent to be
at all tractable.

\subsection{\label{subsec:Convolutional-Neural-Networks}Convolutional Neural
Networks}

Up to this point the network architecture we have discussed have been
fully connected (or dense) feedforward neural networks. However as
discussed previously there is a cost to having many parameters in
the model, in terms of overfitting and in terms of training by calculating
too many gradients. Therefore it is often useful to utilize the structure
of the data in order to reduce the number of parameters used in a
network. Many kinds of data contain a degree of locality, such as
images and text containing local features. The most influential architecture
that takes advantage of this, and also the advancement that took deep
learning from an obscure to a behemoth research field is the convolutional
layer\cite{LeCunn89}. The most common use of convolutional
layers is in 2D image processing tasks, and while eventually the application
relevant to this work is 1D convolutions over text, the concepts are
still the same. 
\begin{figure}
\begin{centering}
\caption{Convolutional Architecture for AlexNet}
\par\end{centering}
\begin{centering}
\includegraphics[scale=0.25]{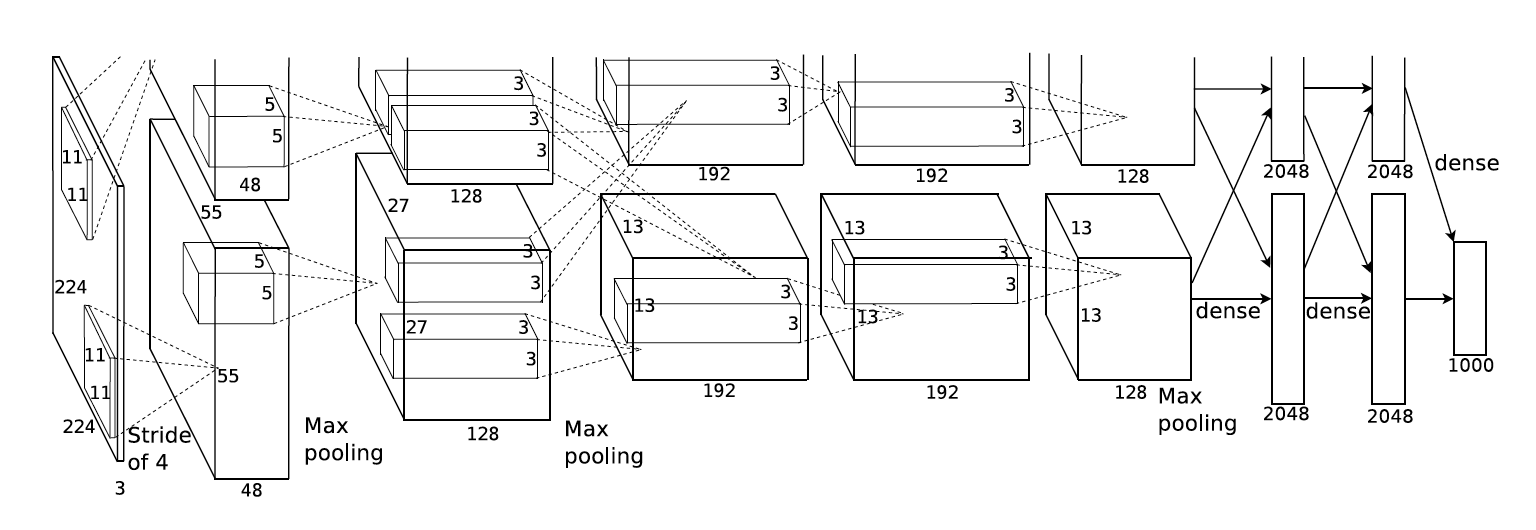}
\par\end{centering}
\centering{}
\end{figure}
 In a convolutional layer a \emph{filter} slides over every position
on the input 2D grid (or more precisely slides over every $n^{th}$
position determined by the stride parameter), and for each position
the filter is dotted with the input. Since the dot product is in some
sense a similarity metric, the output of the convolution can be construed
as the similarity of the local region of the input image and the filter
for every position of the input. For example a filter that has learned
an image like a vertical line, when convolved over the input will
show the locations on the input that seem to contain vertical lines.
With many different filters this can be used to extract many different
local features from the input that do not depend on the specific location
of the feature. These layers are then composed to encode increasingly
abstract information, from lines to shapes to layouts, out of an input,
until eventually the hidden state can be passed to a fully connected
layer and processed in the same way as a feed forward network. The
key benefits of this are that small constant size filters (independent
of input size) have many fewer parameters allowing the model to use
many different filters to be composed to extract a variety of features.
At the same time, the translation independence and locality of the
features makes them more robust. This method also learns how to extract
local features directly from the data, where as previous approaches
had involved using expert knowledge to know what features to look
for. This is what made AlexNet\cite{Krizhevsky:2017:ICD:3098997.3065386}
kick start the current deep learning fervor in 2012 when it vastly
outperformed other models in an image recognition competition.

\subsection{Autoencoders}

\begin{figure}
\begin{centering}
\caption{MNIST Autoencoder\label{fig:MNIST-Autoencoder}}
\par\end{centering}
\begin{centering}
\includegraphics[scale=1.2]{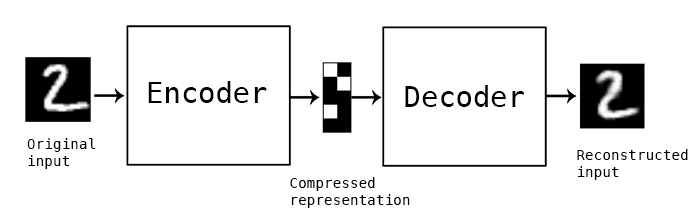}
\par\end{centering}
\centering{}
\end{figure}

The structure of problems this work is concerned with is the class
of latent space models. This is where for a given, probably high dimensional,
dataset we assume that there is some other lower dimensional latent
space that contains all or most of the relevant information of the
data. For instance in the popular MNIST dataset, images are large
28x28 or 784 dimensional vectors, however for the most part the only
useful information can be encoded in a few dimensions representing
the values as well as perhaps a few style attributes. An \emph{autoencoder
}is a deep network that learns this latent representation. The structure
of such a network as shown in figure \ref{fig:MNIST-Autoencoder}
contains three main components. First there is the encoder network,
that transforms the data to some low dimension state, the state itself
at this intermediate layer then encodes all of the information about
the input that the network can access, finally there is the decoder
network that takes this state and then tries to rebuild the original
input. 
 These models are unsupervised, as opposed the more standard supervised
learning paradigm, because it does not require separate labels for
each datum, but rather uses simply the raw data itself as labels in
order to learn just re represent the data as opposed to necessarily
predict something. However this low dimensional representation is
very useful in itself.

Latent space models can be incredibly useful because of the so called
``curse of dimensionality'', where very high dimensional data makes
many kinds of analysis or visualization simply impossible. Building
a low dimensional probabilistic representation of a complex dataset
helps in understanding some of the abstract structure of very complex
high dimensional data. Among these uses is in the ability to easily
sample from a low dimensional space, and then generate good high dimensional
examples from the original data space. It allows for optimization
and visualization techniques that are impossible for very high dimensional
spaces such as Bayesian Optimization\cite{2018arXiv180702811F},
or for tractable comparisons between data that encodes more meaningful
semantic similarity. In particular since the space of possible programs
has a vary high complexity and dimensionality, if a a low dimensional
latent space can be built, there are a huge variety of potential uses
from program synthesis, to encoding program similarity.

\subsection{Natural Language Processing}

Other than computer vision, the most impactful area of application
for deep learning has been natural language processing. In tasks such
as machine translation and automated question answering deep learning
models have clearly outperformed other kinds of models.

The first thing to notice when trying to process text in a machine
learning system is the conversion from the qualitative to the quantitative.
For images this is easy as pixel values are already in a numeric form,
however the input of text needs processing before being able to be
thrown into a neural network. The naive way to input language into
a model is to use a simple one-hot encoding. One-hot vectors are vectors
containing all zeros except for a single value of one at a specified
index. This index is used to represent a single qualitative category,
in this case a word. So a one hot encoding for text would just require
a comprehensive dictionary, and mapping of every word to its corresponding
index. This quickly becomes an issue as vocabularies in natural language
can be very large, and the sparsity of the information for inputs
makes it very difficult to learn patterns from sequences. Therefore
what is commonly used is a word embedding \cite{word2vec}.
A word embedding trains a separate network on a large corpora of text
to predict context words, input a word in one-hot encoding and it
will output a set of predicted adjacent words. After training this
network it then uses the hidden layer state vector whose dimension
is much less than the one-hot vector, as a dense representation for
each word. This process is similar to a more data rich contextual
work level autoencoder. This style of dense representation has been
shown to include a lot of interesting semantic meaning (for instance
the popular example is taking the vector for the work ``king'',
subtracting the vector for ``man'', and adding the vector for ``woman'',
will give a vector closest to the one representing ``queen'').

After encoding words into this vector space, natural language processing
(NLP) becomes a task simply involving one dimensional sequences. While
the most state of the art performant methods use a mechanism called
Attention\cite{AttentionAllYouNeed}\cite{GPT-2},
one dimensional convolutional models also perform very well and require
substantially fewer computational resources\cite{NLP_CNN}.
In this work we will be modeling programing language using these tools
for natural language, as both are essentially the same as far as NLP
tools are concerned. In particular this work will make use of one
dimensional convolutional networks for processing the text of programs.

\subsection{\label{subsec:Graph-Convolutions}Graph Convolutions}

While convolutions have proved to be revolutionary for image processing,
and recurrent networks and attention for language; interest has grown
for structures that can achieve similar results for more complicated
structures than grids or sequences, and generalize to and non-euclidean
topologies\cite{GeometricDeepLearning}\cite{GNN}\cite{graph_survey}.
Because of the added complexity of arbitrary graphs, there is not
one single method that has converged to be the best generalization
of convolution to graph structures. Generally there are spectral methods
that take the whole graph structure, spatial methods that aggregate
local information of nodes based on neighbors, and others such as
Graph Attention Network \cite{GraphAttention} that use mechanisms
like attention from NLP. For the purposes of this work we will focus
on the graph representation learning work from \cite{2018arXiv180204407P}
which uses the Graph Convolution concept from \cite{GraphConvNet}.
\begin{figure}
\begin{centering}
\caption{Graph Convolutional Network}
\par\end{centering}
\begin{centering}
\includegraphics[scale=0.4]{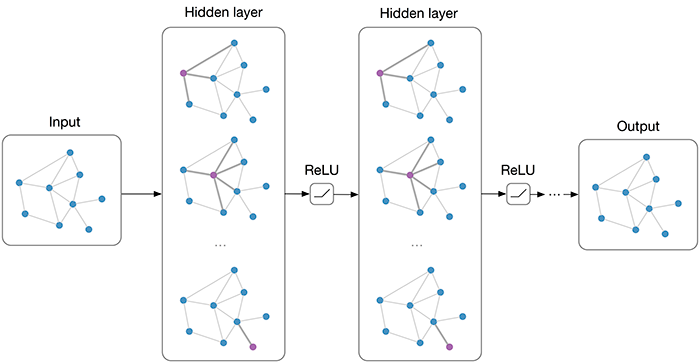}
\par\end{centering}
\centering{}
\end{figure}
 This model uses a graph laplacian to encode local graph information
into successive layers' nodes. Precisely we have a graph input defined
as a vector of node values $X$ and an adjacency matrix $A$. A graph
convolutional layer with learnable weights $W$ on the input graph
$\left\langle X,A\right\rangle $ with activation $\phi$ is defined
\begin{align*}
f_{\phi}(X,A\,|\,W) & =\phi(\tilde{D}^{-1/2}\tilde{A}\tilde{D}^{-1/2}XW)\\
\end{align*}

Where $\tilde{A}=A+I$ to add node self similarity and $\tilde{D_{ii}}=\sum_{j}\tilde{A_{ij}}$
for the local aggregation and then used as $\tilde{D}^{-1/2}$ as
a symmetric normalization of the transformation. The layer outputs
a vector $Z$ which is the same shape as $X$, and successive layers
need to utilize the same structure matrix $A$. However this technique
has still been shown in \cite{2018arXiv180204407P} to be
able to effectively encode structural information purely in the hidden
state $Z$. Essentially what the network learns is how to propagate
information in general between connected nodes.

\subsection{Program Analysis}

These graph layers are relevant to the program synthesis domain because
programs unlike natural language have a formal graph structure underlying
them. The degree to which these graph structures help in auto-encoding
programs in fact is the key question of this work.

There are essentially two kinds of graph information found in programs,
\emph{syntactic} and \emph{semantic} information. Because programs
are written in programming languages, which are formal languages with
a strict grammar, they have syntactic information in the form of the
abstract syntax tree, which is now encodable into a graph convolutional
network. Then there is semantic information, which sadly in general
is an undecidable property thanks to Rice's theorem. However there
are still static program analysis techniques that can unveil useful
if not fully complete information. The one covered in this work is
\emph{control flow analysis}.

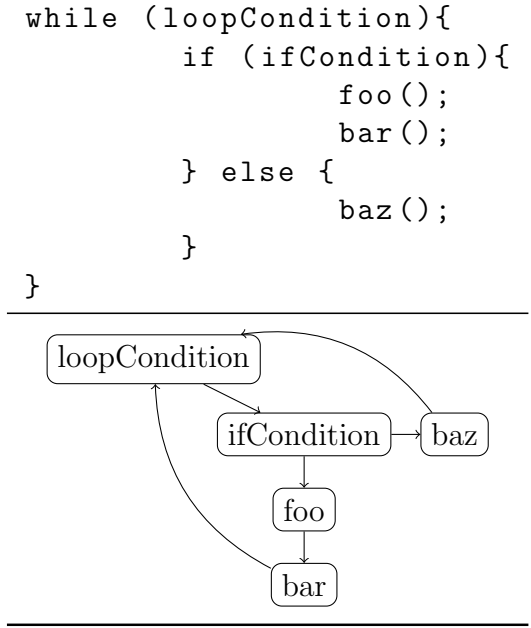
\begin{figure}
\centering{}\caption{Example Program and Control Flow Graph\label{alg:CFG-example}}
\begin{tabular}{c}
\toprule 
\begin{lstlisting}[basicstyle={\ttfamily}]
while (loopCondition){
	if (ifCondition){
		foo();
		bar();
	} else {
		baz();
	}
}
\end{lstlisting}
\tabularnewline
\midrule 
\begin{tikzpicture}[shape=rectangle, rounded corners, align=center]
\node[draw] (0) at (0,0) {loopCondition};
\node[draw] (1) at (2,-1) {ifCondition};
\node[draw] (2) at (2,-2) {foo};
\node[draw] (3) at (2,-3) {bar};
\node[draw] (4) at (4,-1) {baz};
\draw[->] (0) to (1);
\draw[->] (1) to (2);
\draw[->] (2) to (3);
\draw[->,bend left] (3) to (0);
\draw[->] (1) to (4);
\draw[->, bend right] (4) to (0);
\end{tikzpicture}\tabularnewline
\bottomrule
\end{tabular}
\end{figure}

Programs do not always simply execute in the linear line by line order
that they are written in, instead features like loops and conditionals
cause the program counter to jump around from line to line in a nonlinear
fashion. While we cannot predict the exact path that the program will
take, we can build up a graph the models all of the possible paths
a program can take, an example of code and its corresponding graph
can be seen in figure \ref{alg:CFG-example}.
 This graph contains a lot of the overall semantic structure of the
program that is invisible if purely analyzing the text. This graph
is constructed depending on the language, but in general consists
of a set of rules, for each kind of statement how does it potentially
alter the flow of the program. For instance assignments simply pass
the program along from the previous to the next line which would represent
a single directed edge to the next line. While an if statement can
send the program to one of two different points, and so two edges
are added from the if statement, one to each of the potential next
statements. In such a graph any kind of loop or recursion introduces
a cycle, and the final graph is an overestimate of all the possible
traces of a program. This kind of analysis is strictly local and breaks
down when outside procedures are called or if the program modifies
the program counter itself, however in the context of this work we
will ignore such programs and assume all calls halt and bring the
program counter back to to location from which they are called.

Section \ref{sec:Deep-Learning} has introduced the basics of deep
learning and the classes of models that will be used in the rest of
this work. It has also shown how these models might be applied to
programs both as bodies of text and as formal graphs. The section
\ref{sec:Program-Synthesis} on Program Synthesis aims to give a historical
overview of the field of program synthesis and place the potential
applications of this work in a greater context. However it is not
strictly required to understand the methods of this work and so is
more recommended than required reading.

\section{\label{sec:Program-Synthesis}Program Synthesis}

\subsection{History}

Ever since the early days of Computer Science, Program Synthesis has
been an ambitious goal. After all in a field made to automate processes,
among the first tasks a computer scientist might think to automate
is the work she is doing, programming. In the summer of 1957 Alonzo
Church was among the first to formalize this problem, in his case
as building circuits to fulfill certain mathematical properties\cite{church}.
This mathematical and logical framework for the problem persisted
during the early days of artificial intelligence research, with automata
theory approaches and high level programming languages among the takes
on the general problem.

However with the problem being generally considered part of the field
of artificial intelligence, its advancements also fell victim to the
AI winters of the seventies and the nineties. After Churches early
problem statement the most influential framework is that of Mana and
Waldinger in 1980\cite{Manna:1980:DAP:357084.357090}. The
input for this framework is a first order logical formula meant to
specify the properties of a function to synthesize. Then the system
will construct a proof from the formula using such tools as resolution
and induction to build a function that fulfills the specification
as correct by construction. This method could only support very small
functions, and the programming language generated is comparatively
minimal. However the logical deductive strategy more or less set the
direction of the field towards formal logic as the toolset for specification
and construction. Logic is a sensible way to frame program synthesis,
as is a compact way to write exactly the properties one might want
in a program, as well as coming with lots of existing techniques for
manipulating formulae in a very sound manner.

However there is a cost to using logical formulae as the underlying
tool for synthesis, as writing out good logical specifications can
sometimes be just as if not more difficult than the programming we
are trying to automate. So instead of formal specifications, another
approach is to use examples. This approach of programming by examples
is appealing, as anyone regardless of their ability to code or write
formulae can simply do examples of a task that he might want the computer
to automate. A popular way to formulate this approach was that of
Inductive Programming. While Mana and Waldinger introduced a deductive
approach, taking a general principle and deducing the specific program
that fulfilled the principle; inductive programming takes specific
examples and induces a more general program that is consistent with
those examples.

This inductive programming was developed at much the same time as
the field of Machine Learning was being defined, and has a very similar
structure. Leslie Valiant in his 1984 paper for Probably Approximately
Correct (PAC) Learning \cite{Valiant:1984:TL:1968.1972} lays
out the primary framework for machine learning. This approach introduced
computational complexity into the problem, as well as making it a
probabilistic model. This would introduce a split in artificial intelligence
research, as some continued in the formal logical line, while statistical
machine learning developed separately. These two branches came out
of the second AI winter as squarely different fields. Statistical
learning tended to gravitate towards problems in natural language
as opposed to the formal languages of computer programs, while the
field of formal methods came out of the turn of the century with newly
efficient algorithms for boolean satisfiability and later satisfiability
modulo theories that allowed first order logic formulae to be solved
much more efficiently. As a result program synthesis generally remained
in the logical paradigm, and in fact in many departments left AI altogether
in favor of these formal methods and programming languages groups.

Then in 2012 when AlexNet\cite{Krizhevsky:2017:ICD:3098997.3065386}
kickstarted the deep learning revolution in statistical machine learning,
the fields began to come closer together again. Within these well
defined logical frameworks, deep learning models were able to add
a heuristic level improvement in solving strategies, using Neural
Guided Search, NeuroSymbolic Synthesis, or Reinforcement Learning\cite{survey}.
While formal methods remains as the best performant of program synthesis
techniques, further integration of statistical learning has opened
new and exciting areas to explore. This work fits in that tradition,
of deep learning techniques that can be used to potentially augment
existing logical synthesis systems. The following sections will go
into more depth on the workings of existing program synthesis techniques
to give context for application of the deep learning model.

\subsection{Classical Formal Program Synthesis}

\subsubsection{Formal Logic}

The basis for program synthesis engines is formal logic. This section
will give a brief introduction to logic, it will somewhat gloss over
a lot of the nuances and intricacies as this section aims to primarily
just establish a working context and vocabulary for the computational
tools discussed later on. By the end of this section if a reader has
no background in logic beforehand, she should be able to read logical
sentences and be able to understand what they mean at an intuitive
level rather than a strict formal level.

Boolean logic is a system for reasoning about the truth or falsity
of various kinds of statements expressed as logical formulae. The
basic kind of statement is called a logical proposition, and thus
we will begin with propositional logic. A proposition has two essential
components, \emph{variables} and \emph{connectives}. A variable is
any symbol that we determine to represent the truth value of something,
for instance whether it is raining which we can represent in a formula
with the symbol $r$. Connectives combine variables in such a way
that their combination is a logical statement with its own truth value.
For instance we can use the connective \emph{and} $(\text{written as }\wedge)$,
to write the statement, that is is both raining $(r)$ and sunny $(s)$
as $r\wedge s$. This statement has a truth value itself, depending
on the values of the variables that it includes. The set of connectives
that are commonly used are shown in Table \ref{tab:Common-Logical-Connectives}.
\begin{table}
\caption{\label{tab:Common-Logical-Connectives}Common Logical Connectives}

\begin{centering}
\begin{tabular}{|c|c|c|}
\hline 
Symbol (with variable(s)) & Name & Meaning\tabularnewline
\hline 
\hline 
$\neg a$ & Negation & not $a$\tabularnewline
\hline 
$a\wedge b$ & Conjunction & $a$ and $b$\tabularnewline
\hline 
$a\vee b$ & Disjunction & $a$ or $b$\tabularnewline
\hline 
$a\rightarrow b$ & Implication & if $a$ then $b$\tabularnewline
\hline 
\end{tabular}
\par\end{centering}

\end{table}
 The final notational addition are special symbols that represent
something that is always true $(\top)$ or always false $(\bot)$.
A propositional formula can thus be defined recursively: A propositional
formula is either truth ($\top)$, false ($\bot)$, a variable, or
a connective of formula(e).

For example we can write out the proposition, that if it is raining
($r$) then you are wet $w$ or you have an umbrella ($u$) as: 
\begin{equation}
r\rightarrow(w\vee u)\label{eq:(example_prop)}
\end{equation}

This is a valid statement syntactically as it uses the connective
$\rightarrow$ between the variable $r$ and proposition $(w\vee u)$
which is a proposition as it is the connective $\vee$between variables
$w$ and $u$. We can express the truth of this statement, depending
on the values of the variables in something called a truth table.
\begin{table}[ht]
\caption{Truth Table for Equation \ref{eq:(example_prop)}}

\centering{}%
\begin{tabular}{|c|c|c|c|}
\hline 
$r$ & $w$ & $u$ & $r\rightarrow(w\vee u)$\tabularnewline
\hline 
\hline 
$\top$ & $\top$ & $\top$ & $\top$\tabularnewline
\hline 
$\top$ & $\top$ & $\bot$ & $\top$\tabularnewline
\hline 
$\top$ & $\bot$ & $\bot$ & $\bot$\tabularnewline
\hline 
$\bot$ & $\top$ & $\top$ & $\top$\tabularnewline
\hline 
$\bot$ & $\top$ & $\bot$ & $\top$\tabularnewline
\hline 
$\bot$ & $\bot$ & $\top$ & $\top$\tabularnewline
\hline 
$\bot$ & $\bot$ & $\bot$ & $\top$\tabularnewline
\hline 
\end{tabular}
\end{table}
 This table lays out the semantic meaning behind the syntactic logical
sentence. We can use this underlying interpretation to show equivalences
between statements that have the same truth table.
\begin{table}[ht]
\caption{Equivalent Truth Table for Equation \ref{eq:(example_prop)}}

\centering{}%
\begin{tabular}{|c|c|c|c|}
\hline 
$r$ & $w$ & $u$ & $\neg(r\wedge\neg w\wedge\neg u)$\tabularnewline
\hline 
\hline 
$\top$ & $\top$ & $\top$ & $\top$\tabularnewline
\hline 
$\top$ & $\top$ & $\bot$ & $\top$\tabularnewline
\hline 
$\top$ & $\bot$ & $\bot$ & $\bot$\tabularnewline
\hline 
$\bot$ & $\top$ & $\top$ & $\top$\tabularnewline
\hline 
$\bot$ & $\top$ & $\bot$ & $\top$\tabularnewline
\hline 
$\bot$ & $\bot$ & $\top$ & $\top$\tabularnewline
\hline 
$\bot$ & $\bot$ & $\bot$ & $\top$\tabularnewline
\hline 
\end{tabular}
\end{table}
 This allows us to reason about formulae symbolically using equivalent
substitutions (written $\equiv$) such as $(p\vee\neg p)\equiv\top$
or $\neg(p\wedge q)\equiv\neg p\vee\neg q$ and so forth. The truth
table primitive definitions for the common connectives are shown in
Table \ref{tab:Truth-Tables-for-conn}, these connectives are then
composed to form the full set of possible truth tables and thus logical
sentences (in fact you don't even need all of these connectives as
they can be defined in terms of each other, but this set is commonly
used as primitive). 
\begin{table}
\caption{Truth Tables for Common Connectives\label{tab:Truth-Tables-for-conn}}

\centering{}%
\begin{tabular}{|c|c|c|}
\hline 
$p$ & $q$ & $p\wedge q$\tabularnewline
\hline 
\hline 
$\top$ & $\top$ & $\top$\tabularnewline
\hline 
$\top$ & $\bot$ & $\bot$\tabularnewline
\hline 
$\bot$ & $\top$ & $\bot$\tabularnewline
\hline 
$\bot$ & $\bot$ & $\bot$\tabularnewline
\hline 
\end{tabular} %
\begin{tabular}{|c|c|c|}
\hline 
$p$ & $q$ & $p\vee q$\tabularnewline
\hline 
\hline 
$\top$ & $\top$ & $\top$\tabularnewline
\hline 
$\top$ & $\bot$ & $\top$\tabularnewline
\hline 
$\bot$ & $\top$ & $\top$\tabularnewline
\hline 
$\bot$ & $\bot$ & $\bot$\tabularnewline
\hline 
\end{tabular} %
\begin{tabular}{|c|c|c|}
\hline 
$p$ & $q$ & $p\rightarrow q$\tabularnewline
\hline 
\hline 
$\top$ & $\top$ & $\top$\tabularnewline
\hline 
$\top$ & $\bot$ & $\top$\tabularnewline
\hline 
$\bot$ & $\top$ & $\top$\tabularnewline
\hline 
$\bot$ & $\bot$ & $\bot$\tabularnewline
\hline 
\end{tabular} %
\begin{tabular}{|c|c|}
\hline 
$p$ & $\neg p$\tabularnewline
\hline 
\hline 
$\top$ & $\bot$\tabularnewline
\hline 
$\bot$ & $\top$\tabularnewline
\hline 
\end{tabular}
\end{table}

\subsubsection{First Order Logic}

In the propositional logic, every object is logical and has a truth
table, be it a variable or proposition, everything is strictly logical.
However often times we want to be able to reason about nonlogical
objects, and for that we introduce a system called \emph{First Order
Logic}. The key addition in first order logic is the introduction
of quantification over nonlogical variables and boolean predicates
on these variables. A variable that is nonlogical exists over some
domain of discourse, such as the natural numbers, these variables
represent distinct objects in this domain that may have certain characteristics,
but cannot be interpreted as true or false within the formula. In
order to reason about these kinds of objects we define predicates,
which map these abstract objects to the boolean space of true and
false. For instance we can define the $odd$ predicate on the natural
numbers, so $odd(3)$ is true, while $odd(4)$ is false. This predicate
is defined for any natural number in our discourse so we can abstract
a specific number by the symbol $x$ which is a non logical variable,
and then $odd(x)$ becomes a logical predicate depending on the value
of $x$, and since is has a defined boolean value it can be used in
logical formulae.

However in this case $x$ is too abstract to reason about on its own,
and so we \emph{bind} these non logical variables using \emph{quantifiers.}
There are two quantifiers in first order logic: the existential quantifier,
read as ``there exists'' and written as $\exists$, and the universal
quantifier read as ``for all'' and written as $\forall$. These
quantifiers bind non logical variables in formulae, and allow predicates
to have determinate truth value. $odd(x)$ on its own depends of the
value of $x$ which is free since it is an arbitrary variable, however
when we say $\exists x.odd(x)$, or ``there exists an $x$ such that
$odd(x)$ is true'' the truth of the statement is well defined. We
just have to find any specific value for the symbol $x$ within our
discourse, for instance $3$, that satisfies $odd(x)$ to satisfy
the statement; and since we can the statement is true.

This system is clearly very general and powerful, and unsurprisingly
this makes reasoning about arbitrary first order sentences a very
hard problem. It is a generalization of propositional logic, as we
can interpret the logical variables as predicates with no arguments.
However despite it being difficult in general, its expressiveness
makes it useful in many domains.

\subsubsection{Satisfiability}

Now let us return to propositional logic so that we can introduce
the basic problem of satisfiability, this will form the computational
basis for tools that can then be generalized to harder first order
logic. So consider again sentence \ref{eq:(example_prop)}, $r\rightarrow(w\vee u)$.
Whether the proposition is true depends on the values of the variables.
We can call a specific set of assignments to the variables an \emph{interpretation}
or a \emph{model.} While the statement in general is defined by the
truth table, a specific interpretation selects a row from that table,
and thus allows the whole proposition to have a single truth value.
So under the interpretation where the atoms $\{r,u\}$ are true and
$\{w\}$ is false, the proposition can be said to be true. We can
write this out by saying that the interpretation \emph{satisfies}
the formula, since it sets the variables in such a way as to make
the formula true, this is written in our mathematical notation for
a model $M$ and formula $\phi$ as: 
\[
M\models\phi
\]

If a formula has no models that can satisfy it, we say that the formula
is \emph{unsatisfiable}, while if any model satisfies it we say the
formula is \emph{valid.} In general what we want is to be able to
write formulae and then have a computer calculate either a satisfying
interpretation or determine it is unsatisfiable. In general we know
this is a very hard problem, as to check a formula exhaustively would
require checking every row in its truth table, scaling exponentially
with the number of variables. As it turns out this is perhaps the
most well studied problem in computational complexity, and more or
less defines the class of $NP-Complete$. However using lots of very
complicated and fancy algorithms and data structures, this problem
of SAT is one that is in theory very difficult, but in practice somewhat
tractable. So despite a huge explosion in the worst case, much of
the time for small to medium sized problems we can solve the boolean
satisfiability problem.

\subsubsection{Satisfiability Modulo Theories}

Once SAT became a tractable problem in the nineties, the next step
in typical computer science fashion was to use solvers as a backend
for a more expressive language and class of problems. Instead of reasoning
over discrete boolean variables, we want to be able to reason in first
order logic about more complicated objects like numbers or eventually
programs. So we define theories for these domains in first order logic,
and then use these theories in formulae to determine satisfiability,
giving the Satisfiability Modulo Theories (SMT) problem. While most
solvers do not allow quantifiers in the actual formulae they check,
the theories used on the backend will involve first order quantifiers.
This is because in general first order logic is not only NP, but undecidable,
and so the full expressiveness of first order logic is limited to
certain decidable subsets defining theories which can be solved computationally.

The most basic theory is that of equality and uninterpreted functions,
this is sometimes called the empty theory and is used as a base for
other theories. As the name suggests it introduces the equality symbol
($=$)to first order logic which acts over nonlogical values. It also
introduces nonlogical functions, which are called uninterpreted since
they have no definition other than how they are used ($f(x)$ does
not actually calculate some function $f$ but rather defines that
there is a function $f$ defined over the values for $x$). The only
requirement for functions is that they are consistent where if given
the same arguments always have the same output, in formal language
for any function $f$ we have $\forall x_{1},...,x_{n}\forall y_{1},...,y_{n}.(x_{1}=y_{1},...,x_{n}=y_{n})\rightarrow f(x_{1},...,x_{n})=f(y_{1},...,y_{n})$.
This theory allows syntactic properties in whatever universe if being
used to be checked. For instance we can check that two pieces of code
should return the same value. 
\begin{algorithm}
\caption{Example QF\_UF\label{alg:Example-QF_UF}}

\begin{lstlisting}[basicstyle={\ttfamily}]
def foo(x):
	y = x
	z = y
	return z * z

def bar(a):
	return a * a
\end{lstlisting}
\end{algorithm}
 In the code in algorithm \ref{alg:Example-QF_UF}, we can verify
that the two functions return the same value using the theory of equality
and uninterpreted functions. We do this by writing a formula that
asks ``Is there such a input where the outputs of the two functions
differ?''. This gives the following formula
\begin{align*}
(input=x)\wedge(input=a) & \wedge\\
(y=x)\wedge(z=y)\wedge(foo=mul(z,z)) & \wedge\\
(bar=mul(a,a)) & \wedge\\
\neg(bar=foo)
\end{align*}

The formula makes no assumptions over how the $*$ function works,
and yet only assuming the consistency of the function and SMT solver
will quickly show that this formula is unsatisfiable as expected,
showing that the functions return the same value for arbitrary inputs.

There are many theories defined for most SMT systems, like linear
integer arithmetic, linear real arithmetic, bitvectos, and more; these
clearly are useful in understanding programs in a logical manner and
work in very standard ways, having definitions for the relevant sets
of numbers and operations on them $\{+,-,*,\leq,etc...\}$. One of
the more interesting theories that is used is the theory of arrays
or the theory of memories, which is clearly vital if we aim to be
able to verify and eventually synthesize programs. This theory introduces
the nonlogical type of an array, which is essentially a mapping of
indices to values. This involves overloading and allowing the equality
operator to be defined for array types, as well as introducing two
more nonlogical symbols, $read$ and $write$. These definitions form
the axioms of the theory:
\begin{enumerate}
\item Writing a value to an index can then be read from that same index
as that value
\[
\forall i.\;read(write(a,i,v))=v
\]
\item Writing to a different index does not change the value at other indices
\[
\forall i,j.\;\neg(i=j)\rightarrow(read(write(a,i,v),j)=read(a,j))
\]
\item Arrays with the same values at all indices are equal
\[
(\forall i.\;read(a,i)=read(b,i))\rightarrow(a=b)
\]
\end{enumerate}
This in conjunction with a suitable numerical theory generally forms
enough of a background to reason about most programs. Solving formulae
using these theories generally requires solvers to build specific
tactics for how to go about solving them. While some theories are
harder, this division of the more complicated first order logic going
on the side of standard theories allows SMT solvers to be well optimized
for most use cases. Therefore all one needs to do in order to verify
some property of a program is to encode the actions of the program,
be it memory access or arithmetic, into these logical theories; at
which point the SMT solver can check for desired properties. This
on its own is a very useful application of SMT, however we aim to
go one step further and actually synthesize programs that satisfy
a given property.

\subsubsection{Syntax Guided Synthesis}

Now that we have built the background in order to verify properties
in programs, the next step is synthesizing programs that fulfill these
properties. The framework we will introduce is called Counter Example
Guided Inductive Synthesis (CEGIS) and in particular the problem within
that framework of Syntax Guided Synthesis (SyGuS).

The SMT backend we have developed can be used to formalize a certain
property of a program via a formula, and to verify that the property
holds. Therefore the problem of synthesis is actually just a search
problem, over the space of possible programs. This general paradigm
is expressed in the CEGIS framework. 
\begin{figure}
\caption{CEGIS Framework}

\begin{adjustbox}{center}
\begin{tikzpicture}[shape=rectangle, rounded corners, align=center]
\node[draw] (0) at (0,0) {Synthesizer};
\node[draw] (1) at (-2.5,-1) {Trial Program, $P$};
\node[draw] (2) at (0,-2) {Verifier};
\node[draw] (3) at (2.5,-1) {SAT: Counterexample};
\node[draw] (4) at (5,-2) {UNSAT: $P$ is a solution};
\node[draw] (5) at (4,0) {No Solution};
\node[draw] (6) at (-6,-2) {Formula, $\Phi$};
\draw[->] (0) to (1);
\draw[->] (1) to (2);
\draw[->] (2) to (3);
\draw[->] (3) to (0);
\draw[->] (2) to (4);
\draw[->] (0) to (5);
\draw[->] (6) to (2);
\end{tikzpicture}
\end{adjustbox}

\end{figure}
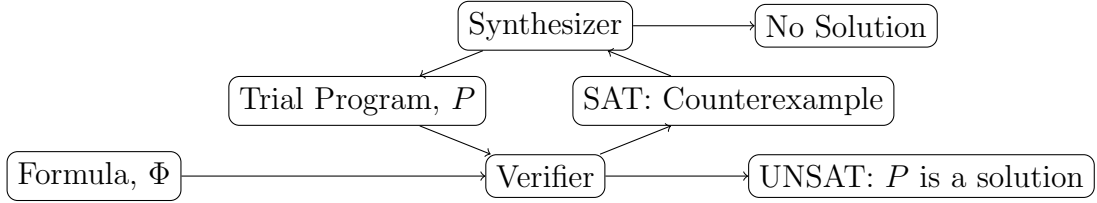
 In this framework the only hole left to fill in is the synthesizer,
which has a lot of freedom in how it works. All the synthesizer now
has to do is try as many different programs as it likes, and every
time it gets it wrong it can potentially learn from the generated
counterexample. The SyGuS problem adds one additional component of
structure to the system, by stating the the synthesizer can only generate
programs that comply with a specific grammar. This helps limit the
search space from the whole universe of text to a small set of syntactically
correct programs, so the grammar ensures syntax while SMT solver ensures
semantics.

While this framework allows for potentially very advanced synthesizers,
however currently the state of the art, as determined by the SyGuS
competition\cite{sygus-org}, utilizes essentially a well
optimized enumerative solver, trying programs from the syntax exhaustively.
This goes to show both how hard the problem is, that there is nothing
that obviously outperforms a somewhat naive solution, and that there
is still tremendous potential in developing synthesis tools. The next
section will explore exciting new directions that are being researched
that can fit into this overall CEGIS framework.

\subsection{Neural Program Synthesis}

Historically program synthesis has been squarely framed in terms of
formal logic. However recent development in the CEGIS framework has
opened the door for new kinds of approaches to program synthesis that
do not involve logic programming or formal methods, as those components
are abstracted away by the framework and computed by a separate SMT
solver. This opens the door deep learning researchers to latch themselves
onto yet another branch o computer science. Here there are two popular
ways to integrate modern machine learning into this problem. Either
building a model to learn to synthesize programs all on its own, in
a reinforcement learning type setting, or using the immense amount
of data generated from trying different programs to build models that
can act as heuristics in other synthesizers. The latter is currently
the most successful, and is in use by Microsoft Research and others
and is a very active area of research\cite{2018arXiv180902840Z}\cite{Lee:2018:ASP:3192366.3192410}\cite{2018arXiv180401186K}\cite{2018arXiv180204335P}\cite{Feng:2018:PSU:3192366.3192382}.

At the same time work has been done in representing the text of programs
in embedded form more suitable for deep learning techniques\cite{code2vec}.
This work follows that path and aims to build a semantically meaningful
encoding of program text in a probabilistic latent space. This could
fit into a CEGIS style solver by being able to search a lower dimensional
and more semantically meaningful space of programs.

\section{\label{sec:Similar-Work}Similar Work}

There have been a few similar works in program embeddings and it is
a very new and active area of research. In \cite{dynamicNeuralPorgramEmbedding}
embedding is assisted by semantic information in the form of live
program traces, using recurrent neural networks to encode information
from traces. This differs from the goal of this work which is to encode
purely statically, however it does show the value of semantic information
in the effectiveness of program embedding. More recently in 2019 there
have been two interesting works worth comparing to. The SemCluster
\cite{SemCluster} approach clusters similar algorithms using
a more formal approach, classifying code based on partitions of the
input space of the program into equivalence classes. The code2vec
paper\cite{code2vec} is the closest to this work, using a
neural model to encode programs into a vector space, using both text
and the abstract syntax tree. The primary difference is first in the
neural architecture being used, code2vec uses an attention mechanism
while this work uses convolutional and graph convolutional layers.
The other primary novelty in this work is it is the first the analyze
the effectiveness of these models not just on syntactic graphs, but
on static semantic analysis in the form of control flow diagrams.

\chapter{Methods}

\section{Goals}

\begin{figure}
\begin{centering}
\caption{\label{fig:General-Embedding-Pipeline}General Embedding Pipeline}
\par\end{centering}
\centering{}\begin{tikzpicture}[shape=rectangle, rounded corners, align=center]
\node[draw] (0) at (0,0) {Program Text};
\node[draw] (1) at (0,-1) {Preprocessing to Input Representation};
\node[draw] (2) at (0,-2) {Encoder Network};
\node[draw] (3) at (0,-3) {Latent Representation};
\node[draw] (4) at (0,-4) {Decoder Network};
\node[draw] (5) at (0,-5) {Output Reconstructed Representation};
\node[draw] (6) at (0,-6) {Reconstructed Estimated Program Text};
\draw[->] (0) to (1);
\draw[->] (1) to (2);
\draw[->] (2) to (3);
\draw[->] (3) to (4);
\draw[->] (4) to (5);
\draw[->] (5) to (6);

\end{tikzpicture}
\end{figure}
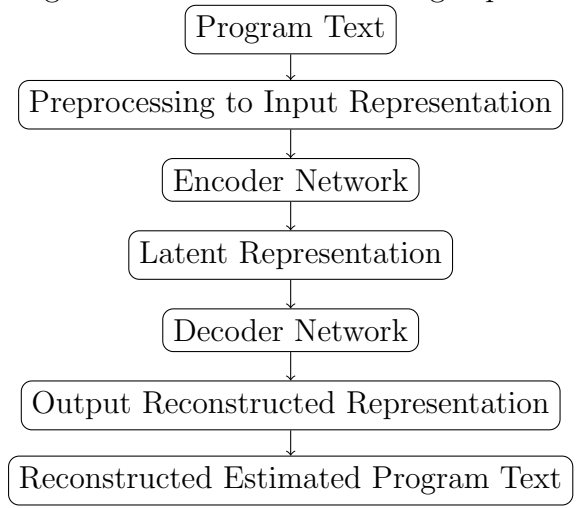

Since the impetus for this study is in program synthesis, the goal
must be to study the potential for these different models to generate
programs. However full program synthesis tools in this case would
add needless complication to the more restricted study of this work,
therefore the program embedding goal was chosen. The task of embedding
involves encoding as much information about a program text as possible,
and then the ability to regenerate a program from that information
as correctly as possible. This covers the core competency of program
understanding and generation in the program synthesis world, while
reducing the need for more complicated frameworks. The general pipeline
will be like in figure \ref{fig:General-Embedding-Pipeline}, where
raw code will be processed then encoded. 

Therefore within the framework of embedding there are a few key evaluation
metrics to use. Firstly there is the actual loss function comparing
a reconstruction and original, in this case the Cross Entropy. There
is also the simple accuracy, the rate at which the reconstruction
correctly generates the same tokens as the original programs. There
are also higher level ways to evaluate the effectiveness of the embedding,
such as procedure name comparison. This is where programs are embedded
anonymously without names, and then programs close to each-other in
the embedding space are compared to see if they have real semantic
similarity, which would be a good sign of the model embedding meaningful
information about the programs.

These comparisons will be done on a few classes of models. Firstly
are the Graph Convolutional Models. In these models it is easy to
compare static analysis to a control, this is done by using two equally
sized graph convolutional networks, and training one with a control
flow graph as the adjacency matrix, and comparing with a linear control
group that uses an adjacency matrix representing just a linear flow
from token to token which is the same assumption made in normal linguistic
models. This allows a truly apples to apples comparison as the networks
will have identical size and design, and their only difference will
be the degree to which they have been provided nontrivial analysis
graph representations. The hypothesis going in to this experiment
is that models using analysis graphs should significantly outperform
naive models. The success or failure of this hypothesis will then
be tested using the above mentioned metrics.

\section{Design}

\subsection{Dataset}

As with all machine learning problems the single most important component
is the data. For this work two different datasets utilizing different
programing languages were considered, initially the Natural Program
Synthesis Dataset or NAPS\cite{NAPS}, and then a dataset
of Java open source code\cite{code2vec}$^{,}$ \cite{allamanis2016convolutional}.
While the primary insights of this work used the Java data, both provided
useful information and will be described here.

The NAPS dataset aims to build a large library of 'competitive programming'
solutions in order to provide a basis of 'algorithmic problems'. Competitive
programming competitions, in this case from the website codeforces,
emulate programming problems of the kind found in technical interviews.
This involves traversing different data structures, dynamic programming,
and the like. This is appealing for the program synthesis space as
programs are small in the number of lines used, and strictly limited
in scope consisting of a single function designed to do a single task.
And while the problems are simple in one aspect, they also have among
the most complex semantics, involving subtle algorithms and techniques,
of other program synthesis data. This should in principle allow models
to learn complex fundamentals of algorithmic thinking while being
able to remain a set of unique and distant data, as opposed to natural
code with complex interdependencies that does not generally lend itself
to independent and identically distributed code blocks.

This dataset scrapes solutions from specific codeforces competitions,
limiting itself to easy and intermediate level problems. However these
competitions allow a variety of programming languages to be used,
so in order to aggregate as much data into the same form as possible,
compatible solutions are converted into a custom domain specific language
called UAST. This DSL aims to maintain the overall readability of
programs while getting rid of runtimes or compilation steps. It is
converted to from the range of languages found in the codeforces competitions,
Java, C++, C\#, Pascal, and Python. The preprocessing done also aims
to anonymize the problem specifications, and variables are all names
simply in order of appearance ($var0$, $var1$, etc... ). NAPS currently
consists of 16410 training examples and 485 test examples as split
by the original authors.

The NAPS dataset is a very exciting and ambitious new dataset, which
is why it was initially chosen for this work. However ultimately the
difficulty associated with the dataset outweighed the gains for the
goals fo this project. Firstly the dataset is somewhat small by many
deep learning standards, and initial explorations found most well
performing models to be susceptible to overfitting because of this.
In addition the actual programs, while not large, are highly nuanced
and complex, making the task of learning their representations very
difficult, which made showing differences between the different models
difficult to ascertain. And so since the essential goal of this work
is not to build a state of the art model per se, but rather to make
a strong empirical comparison between equivalent models, it was determined
that a larger and easier dataset would be more appropriate.

Therefore the dataset that was ultimately used is the Java open source
data from \cite{code2vec}$^{,}$ \cite{allamanis2016convolutional}.
As described in their paper \cite{allamanis2016convolutional},
11 open source Java projects from GitHub were clones. The most popular
projects were found by taking the sum of the z-scores of the number
of watchers and forks of each project, using GHTorrent\cite{Gousios:2012:GGD:2664446.2664449}.
The top 11 projects were chosen that contained more than 10MB of source
code files each. These projects have thousands of forks and stars,
being widely known among software developers. The projects along with
short descriptions are shown in Table \ref{tab:Open-Source-Projects}.
Using this procedure a mature, large, and diverse corpus of real source
code is selected. 
\begin{table}
\begin{centering}
\caption{\label{tab:Open-Source-Projects}Open Source Projects Used}
\par\end{centering}
\begin{centering}
\begin{tabular}{lll}
\toprule 
Project Name & Git SHA & Description\tabularnewline
\midrule
\midrule 
cassandra & 53e370f & Distributed Database\tabularnewline
\midrule 
elasticsearch & 485915b & REST Search Engine\tabularnewline
\midrule 
gradle & 8263603 & Build System\tabularnewline
\midrule 
hadoop-common & 42a61a4 & Map-Reduce Framework\tabularnewline
\midrule 
hibernate-orm & e65a883 & Object/Relational Mapping\tabularnewline
\midrule 
intellij-community & d36c0c1 & IDE\tabularnewline
\midrule 
liferay-portal & 39037ca & Portal Framework\tabularnewline
\midrule 
presto & 4311896 & Distributed SQL query engine\tabularnewline
\midrule 
spring-framework & 826a00a & Application Framework\tabularnewline
\midrule 
wildfly & c324eaa & Application Server\tabularnewline
\bottomrule
\end{tabular}
\par\end{centering}
\end{table}

While using open source repositories has the benefit of a very large
corpus, there is the cost of having code that is natural, and therefore
being formatted in larger classes and not the small self contained
blocks from NAPS. In order to make the size of each datum for our
analysis tractable we then extracted methods from the raw Java classes
and performed our analysis on these processed methods. The exact procedure
used to process the methods will be further explained in section \ref{subsec:Feature-Engineering}.
Despite the difficulty in using purely natural code, there are also
benefits in terms of the applicability of the results of this work.
Since the ultimate goal of program synthesis is in generating not
just code, but ideally readable code or human-like code, the ability
of models to represent code not just in a vacuum but in the real applications
that need to be built is a more valuable trait to evaluate. These
natural code methods are also in general much simpler than the complex
puzzles found in NAPS, which means that the models are better able
to represent the code and allowing for a more effective comparison
between models.

\subsection{Feature Engineering\label{subsec:Feature-Engineering}}

Since the Java data consists of a large directory structure of raw
Java files, various preprocessing had to be performed in order to
extract the features and representations used in the experiment. This
consists of three main phases, extracting methods from the large library
of Java class files, generating the sequence of processed tokens used
to represent the text of the methods, and the static analysis phases
generating a control flow graph.

Since we are dealing with raw java code from real programmers, there
is some variance in certain style conventions. In order to reduce
variance and make different analysis steps easier, the first transformation
performed on the code is some automatic formatting through Eclipse.
This ensured consistent use of brackets in all situations where they
are optional and other automatic formatting steps to increase verbosity.

The next step in building a dataset suitable for deep learning models
is extracting the relevant text sequences. As stated before, only
methods are considered as they represent the best atomic units of
computation that we want to represent. This is done by traversing
the tree directory structure of the different repositories in the
dataset, and for each class parsing the Java code. The classes do
not need to be fully compiled since we do not really care about running
the code, only extracting the relevant components. The abstract syntax
tree is then traversed, where all method declaration nodes are processed.
Then any method that is abstract, or a constructor are ignored. This
is because we are trying to consider actual computational blocks,
and obviously abstract methods contain no code and constructors only
have meaning in so far as they instantiate the class, which for the
purposes of this work are ignored, and so constructors are also not
useful. This set of method declarations and bodies is then passed
to the next processing step.

\begin{figure}
\noindent \caption{\label{fig:Vocabulary-Frequencies}Vocabulary Frequencies}
\centerline{\includegraphics[scale=0.4]{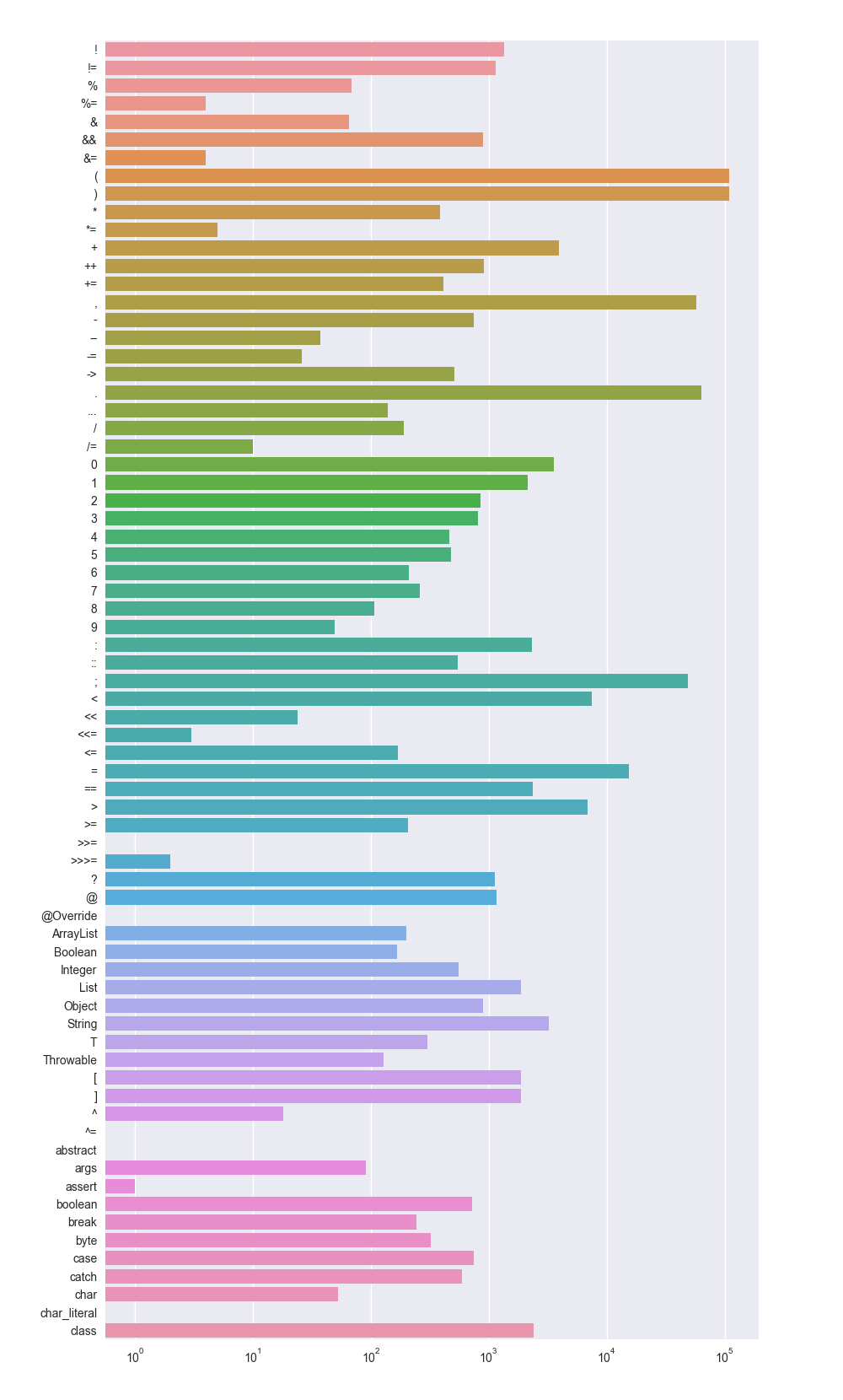}\includegraphics[scale=0.4]{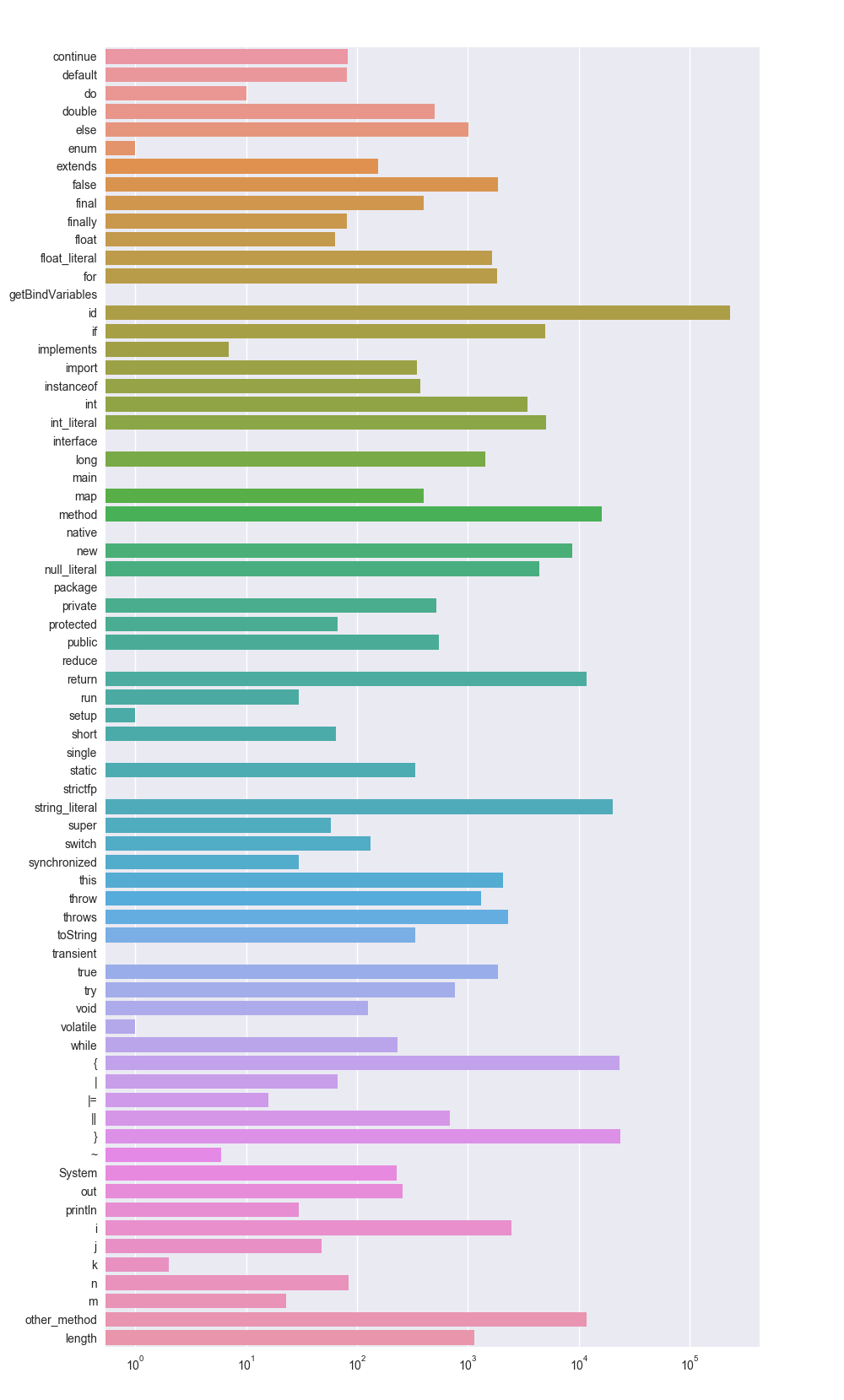}}
\end{figure}

After extracting the blocks of text that are set to be analyzed, they
need to be converted into a format that is workable within a machine
learning setting. This involves two procedures, first is anonymization,
and the second is numericalization. First is anonymization. Since
point of this work is for the most part concerned with the structure
of computer programs, not variable naming conventions we want to make
identifier names anonymous. Also, since ultimately the full corpus
of processed method text needs to be contained within a fixed sized
vocabulary it would be impractical to include every variable name
individually. This also has the effect where the model does not know
the actual name of each method, and therefore allows for evaluation
criteria involving name prediction and similarity. Therefore identifiers
with unique names are replaced in text with the $id$ token. Relationships
between which $id$ tokens refer to the same variable are then saved
as well for potential use in certain models, however this is not strictly
necessary for the techniques explored in this work. The anonymization
however is not universal. A few common variable names that could be
useful to differentiate are maintained, such as $i$ and $j$ (see
figure \ref{fig:Vocabulary-Frequencies}). In addition to variable
names, method names are important to differentiate. Recursive invocations
of the method being processed (in addition to the actual name in the
declaration) are replaced with the $method$ token, while invocations
of other methods are replaced with the $other\text{\_}method$ token.
This process thus transforms the specific text of each method into
a general and universal vocabulary and format, and thus allows for
patterns within the actual structure of methods to be learned.

Next is the numericalization process. Since the neural networks require
numeric inputs as opposed to strings the sequence of tokens must be
converted to a string of vectors. The first step of this process is
to develop a vocabulary for the whole corpus of the dataset. Once
the entire set of used tokens is known, we can represent each token
by a unique number being the index of their location within that vocabulary.
That index will then be converted to a one-hot encoding vector for
each token, this makes each qualitatively different token independent
from each other without the false quantitate information of just the
raw index number. In most natural language processing tasks this process
goes one step further and the one-hot encoded word representation
is converted to a dense embedding, however in this case there are
actually not that many different tokens that need to be considered.
Due to the anonymization process, the size of the vocabulary is tractably
small and thus the network is able to function using the raw encoding
of the tokens. This helps reduce the variance in our results as there
does not need to be a training process to build deep embeddings for
each token, and the only training process in the whole work will be
the direct comparison of models. The full set of 143 tokens in the
vocabulary and their frequencies is shown in table \ref{fig:Vocabulary-Frequencies}.
 This vocabulary includes the expected syntax of java, as well as
some of the specially included identifiers (like common type names
and variable names), and some common literals (like single digit integers)
and additional terms for other kinds of literals. This completes the
process so that the raw java is converted into sequences of tensors
that can be inputted into deep learning models. The final preprocessing
step is the static analysis, being the primary experimental factor
in this work.

It is important to note that this process abstracts away a huge amount
of information. In fact the resultant text does not super closely
resemble useable code, but rather is closer to a boilerplate pseudocode
template. Such a template could still be useful as an intermediary,
where formal methods can fill in specific identifiers and so forth,
and the template only includes high level structural information.
Of course this means that the learning task being trained for is substantially
limited, and as is shown later, this does have a significant effect
on the result. However this is in fact a necessary evil of deep learning
or pure statistical language model techniques, since these models
require a fixed vocabulary that would require a certain amount of
abstraction no matter what. Therefore this analysis is still meaningful
in so far as it is comparing where and how deep learning models perform
when augmented with formal data. Yet we must still be cognizant of
the limitations of this kind of formulation.

\subsection{Static Analysis Step}

\begin{figure}[ht]
\begin{centering}
\caption{\label{fig:Preprocesing-Pipeline}Preprocessing Pipeline}
\par\end{centering}
\centering{}\begin{tikzpicture}[shape=rectangle, rounded corners, align=center]
\node[draw] (0) at (3.5,0) {Program Text};
\node[draw] (1) at (3.5,-1) {Parsing};
\node[draw] (2) at (0,-2) {Anonymization and Numericalization};
\node[draw] (3) at (0,-3) {Tensor Sequence};
\node[draw] (4) at (7,-2) {Static Analysis};
\node[draw] (5) at (7,-3) {Control Flow Graph Adjacency Matrix};
\draw[->] (0) to (1);
\draw[->] (1) to (2);
\draw[->] (2) to (3);
\draw[->] (1) to (4);
\draw[->] (4) to (5);

\end{tikzpicture}
\end{figure}
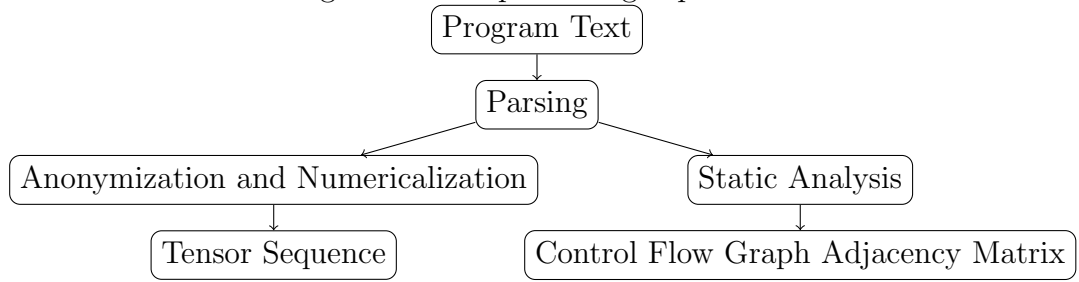

The final and most important step in the processing of the input data
for this work is the static analysis step. In figure \ref{fig:Preprocesing-Pipeline}
the preprocessing pipeline is shown. In it can be seen the two forms
of input data that are generated, the raw code sequence tensors, and
the control flow graph adjacency matrix. Later models will be compared
when given inputs of just the code sequence, or the code sequence
with the control flow matrix and the difference in performance will
be analyzed. 

The static analysis being performed here is among the most simple,
and making the fewest assumptions about the code as possible. In this
analysis it is assumed all procedure calls halt and return back to
where they were invoked. The flow of the program is modeled from token
to token in order to match the dimension of the outputted matrix with
the token sequence. While it is very possible to reduce the size of
this graph to more basic blocks without a loss of generality, the
formatting gain in this work is more valuable as it allows a cleaner
way to compare the addition or subtraction of this information within
the context of graph neural networks. In this vein there are a huge
number of optimizations and other choices that could have been made
in building this analysis system that were omitted in order to provide
the most basic, comparable, baseline result possible.

Control flow from token to token is modeled through an adjacency matrix
of size $n\times n$ where $n$ is the length of the sequence for
a given method. Values of $1$ at location $i,j$ represent that there
may exist a flow of the program from token $i$ to token $j$. Of
course since this is a simple analysis and in general information
about whether such a transition will take place is undecidable this
only provides a loose model of the structure of the method. However
since this analysis never uses this graph in any mathematical modeling,
but rather tries to learn patterns statistically from how different
programs are structured this is not an issue.

For this analysis there are 7 special tokens which alter the flow
from token to token, $if,\;else,\;do,\;while,\;for,\;return,\;\text{and }method.$
All other tokens are assumed to simply pass control linearly along
to the next token. These other special tokens however edit the control
flow graph (specific implementation details for how the sequence is
actually processed are in section \ref{sec:Implementation}). When
encountering an $if$ token there are two edges added. The graph will
continue linearly through the conditional statement, however once
reaching the actual enclosed code block in brackets there are two
edges, one continuing flow into the block and one skipping the block
entirely and connecting the two brackets in the adjacency matrix.
The $else$ tokens work in much the same way, without the conditional
statement ($else\;if\,$ statements are therefore just modeled as
equivalent to $if$ statements). Clearly this is an over-approximation
of real possible paths, since for instance in a simple $if...else...$
combination the graph makes it possible to traverse without entering
either block. However for the reasons explained earlier, this is not
so much of a problem in representing the overall structure of the
method in a way to be learned statistically as opposed to formally,
given the inherent restrictions of static analysis.

The looping tokens of $\ do,\;while,\;\text{and }for$ all act essentially
the same in this analysis. After the condition (at the begging of
the loop for $while$ and $for$ and at the end of the loop for $do$)
two edges are added to the graph, one representing continuing on with
the rest of the program and one connecting back to the beginning of
the loop, introducing a cycle into the control flow graph. The other
looping mechanism considered in the analysis step is simple recursion.
When encountering the $method$ token, which represents the name of
the current method, an edge is added from the end of the invocation
to the beginning of the sequence. An edge is also added continuing
on linearly with the program. This therefore implies that the recursion
is not infinite and that eventually the call will terminate.

The final special token is the $return$ token. When encountering
this token, the flow graph will continue processing up until the next
semicolon for the end of the full return statement, (this allows things
like recursive calls within a return statement), and then add an edge
going straight to the end of the method, and not add any other possible
flows.

\subsection{Other Formats}

There are many different forms of static analysis that could result
in other interesting graphs for this kind of experiment. This section
will discuss some of those alternatives and why they were not selected.

The most obvious structure that was omitted from this work is the
abstract syntax tree. The syntactic information of the AST also fits
nicely within a graph structure, is easy to parse out, seems to be
very representative of the overall code structure, and is used in
other works on program embedding (albeit in very different ways).
There are two main reasons that the AST was not part of the comparisons
in this work: non-matching structure and redundant information. When
parsing a method, the abstract syntax tree will contain nodes not
just from the terminal tokens, but also representing different abstract
nodes from non-terminals in the grammar. This causes the structure
of the AST graph to not match the raw code sequence. This means that
the technique in this work of Graph Convolutional networks over the
code sequence, with varying adjacency matrices could not work as the
resultant graph and matrix for the AST would not 'fit' for analyzing
the raw code. This would mean for the comparisons in the work either
some information about the size of the AST graph would be available
for the control models, or else the comparison would not be using
equivalent data. This complication did not appear to be worthwhile
for this work, as the only information gained form the AST is syntactic,
lacking the semantic analysis information that is the heart of this
comparison. This is to say that the information from the AST is redundant
from what the model could learn just from the raw text.

The other kind of alternative is different kinds fo static analysis.
Beyond a control flow graph there are other kinds of graphs that could
be generated statically from source code, from liveness analysis to
last-write analysis. However almost all of these analyses are actually
predicated on control flow information, and control flow information
better portrays the large scale structure of the program. Therefore
if control flow provides the best first order approximation of the
program among static analysis techniques, and is required for more
advanced analyses, then for the purposes of having the most clear
baseline comparison it is the obvious choice for this work. Other
frameworks may be extended as future work, however for an initial
exploration analyzing the effectiveness of this class of techniques,
it is better to be more limited in scope.

\subsection{Neural Architectures}

After the data has been processed to a form acceptable for deep learning
models, these models can be trained. The primary comparison of this
work utilizes the Graph Convolutional Network architecture from \cite{GraphConvNet}
which was explained generally in section \ref{subsec:Graph-Convolutions}.

\begin{figure}[ht]
\begin{centering}
\caption{\label{fig:Training-Model}Model Comparison}
\par\end{centering}
\centering{}\begin{tikzpicture}[shape=rectangle, rounded corners, align=center]
\node[draw] (0) at (0,0) {Program Text};
\node[draw] (1) at (0,-1) {Parsing};
\node[draw] (2) at (-3.5,-2) {Anonymization and Numericalization};
\node[draw] (3) at (-3.5,-3) {Tensor Sequence};
\node[draw] (4) at (3.5,-2) {Static Analysis};
\node[draw] (5) at (3.5,-3) {Control Flow Graph Adjacency Matrix};
\node[draw,scale=0.5] (6) at (-6,-3.75) {Linear Flow Adjacency Matrix};
\node[draw,scale=0.5] (7) at (-6,-4.25) {Identidy Adjacency Matrix};
\node[draw] (8) at (5,-5) {Sequence Graph Model};
\node[draw] (9) at (0,-5) {Linear Graph Model};
\node[draw] (10) at (-5,-5) {Naive Graph Model};
\node[draw] (11) at (0,-6) {Tensor Sequence (Reconstruction)};
\node[draw] (12) at (0,-7) {Program Text (Reconstruction)};
\draw[->] (0) to (1);
\draw[->] (1) to (2);
\draw[->] (2) to (3);
\draw[->] (1) to (4);
\draw[->] (4) to (5);
\draw[->] (5) to (8);
\draw[->] (3) to (8);
\draw[->] (6) to (9);
\draw[->] (3) to (9);
\draw[->] (7) to (10);
\draw[->] (3) to (10);
\draw[->] (8) to (11);
\draw[->] (9) to (11);
\draw[->] (10) to (11);
\draw[->] (11) to (12);
\end{tikzpicture}
\end{figure}
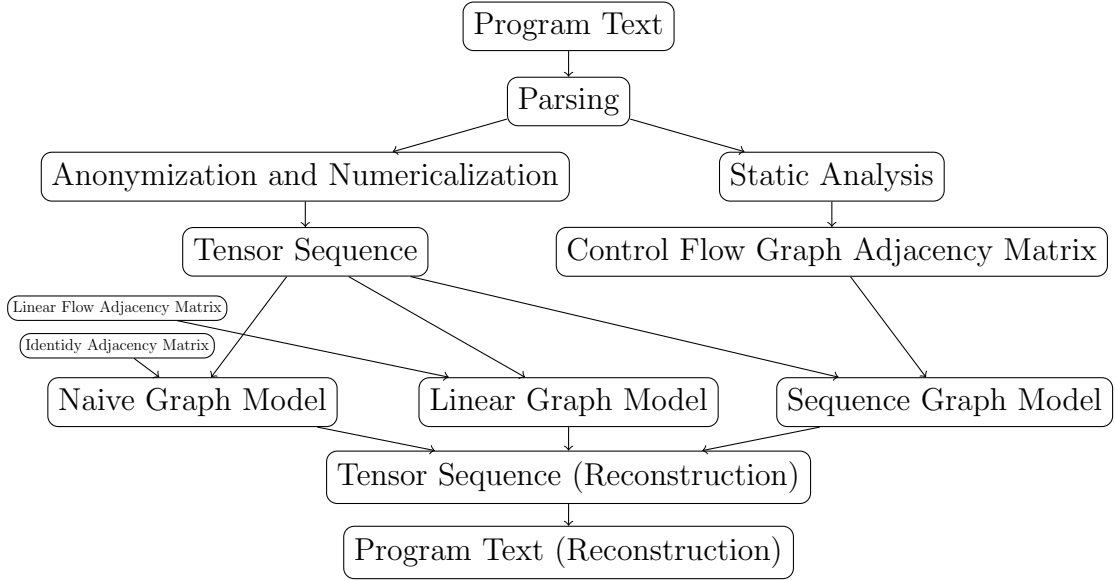

The processed data will be used to train a Graph Convolutional Autoencoder.
This requires two inputs, the code sequence and adjacency matrix.
The model then encodes these inputs into a smaller dimensional latent
space, and then decodes that space back to produce an output sequence.
Because it is an autoencoder, this output sequence is set to be the
same as the input sequence, thereby measuring how well the latent
space can represent and reconstruct the program text. This work compares
completely equivalently specified Graph Convolutional Autoencoders,
with the only difference being the adjacency matrix that they are
given during training. Figure \ref{fig:Training-Model} showcases
the pipeline on how the different models are trained. 
 There are three different situations this work will compare within
the Graph Convolutional framework. Those are using the Control Flow
or Sequence graph input, using a linear graph input, and a naive input.

\begin{figure}
\centering{}\caption{Example Program and Different Graph Inputs\label{fig:Diff-Graphs}}
\begin{tabular}{cc}
\toprule 
Example Program & 
\begin{lstlisting}[basicstyle={\ttfamily}]
while (loopCondition){
	if (ifCondition){
		foo();
		bar();
	} else {
		baz();
	}
}
\end{lstlisting}
\tabularnewline
\midrule 
Sequence Graph & \begin{tikzpicture}[shape=rectangle, rounded corners, align=center]
\node[draw] (0) at (0,0) {loopCondition};
\node[draw] (1) at (2,-1) {ifCondition};
\node[draw] (2) at (2,-2) {foo};
\node[draw] (3) at (2,-3) {bar};
\node[draw] (4) at (4,-1) {baz};
\draw[->] (0) to (1);
\draw[->] (1) to (2);
\draw[->] (2) to (3);
\draw[->,bend left] (3) to (0);
\draw[->] (1) to (4);
\draw[->, bend right] (4) to (0);
\end{tikzpicture}\tabularnewline
\midrule 
Linear Graph & \begin{tikzpicture}[shape=rectangle, rounded corners, align=center]
\node[draw] (0) at (0,0) {loopCondition};
\node[draw] (1) at (2,-1) {ifCondition};
\node[draw] (2) at (2,-2) {foo};
\node[draw] (3) at (2,-3) {bar};
\node[draw] (4) at (4,-1) {baz};
\draw[->] (0) to (1);
\draw[->] (1) to (2);
\draw[->] (2) to (3);
\draw[->] (3) to (4);
\end{tikzpicture}\tabularnewline
\midrule 
Naive Graph & \begin{tikzpicture}[shape=rectangle, rounded corners, align=center]
\node[draw] (0) at (0,0) {loopCondition};
\node[draw] (1) at (2,-1) {ifCondition};
\node[draw] (2) at (2,-2) {foo};
\node[draw] (3) at (2,-3) {bar};
\node[draw] (4) at (4,-1) {baz};
\end{tikzpicture}\tabularnewline
\bottomrule
\end{tabular}
\end{figure}
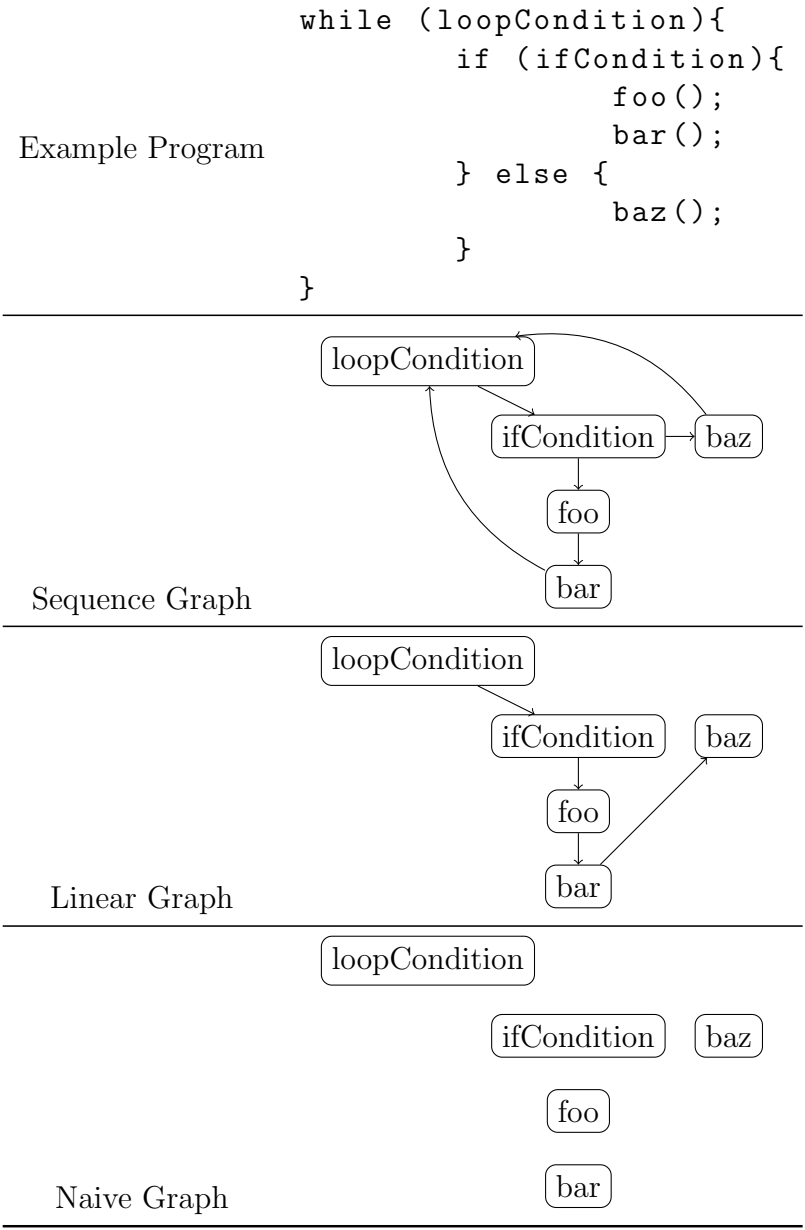

The sequence graph works as described earlier, representing the sequence/flow
of the program from token to token. The linear graph input makes the
linguistic assumption of a simple linear sequence of tokens, this
is to say the adjacency matrix given to the model consists just of
nodes pointing from each token to the next in a linear fashion. Finally
the naive baseline provides no graph information whatsoever to the
model (which in effect inputs the identity matrix due to the workings
graph convolution \cite{GraphConvNet}) which essentially
reduces the problem to a very simple token encoding problem. Therefore
using identical models, these three pipelines can test how useful
the additional contextual information provided by the graphs are.
The initial hypothesis being that the sequence graph will perform
better than the linear which will perform better than the naive. This
hypothesis will then be tested by training the models through the
described pipeline and measuring and comparing the performance difference.

\section{\label{sec:Implementation}Implementation}

This section will now provide a more detailed explanation of the implementation
of the design set above, in addition the full source code can be found
at https://gitlab.doc.ic.ac.uk/aw2318/msc-individual-project/ and
is also included alongside this report in the project submission.
Python was chosen as the standard environment for deep learning research,
and PyTorch was chosen as the deep learning framework. PyTorch was
chosen over other frameworks like Tensorflow due to readability, consistency
with Imperial College coursework practice, and available open source
implementations. The most valuable of these reasons is the open source
implementations. The Graph Convolutional Network paper has made their
code open source and available on Github (https://github.com/tkipf/pygcn).
By using this code as the basic units for the Graph Convolutional
Autoencoders this work is able to ensure greater reproducibility,
and avoid potential bugs. Algorithm \ref{alg:Graph-Convolutional-Layer}
shows the specific code defining the GraphConvolution layer which
is essentially unchanged from the original paper. Algorithm \ref{alg:Graph-Convolutional-Autoencoder}
shows the code for the full autoencoder. 
\begin{algorithm}
\caption{\label{alg:Graph-Convolutional-Layer}Graph Convolutional Layer}

\begin{lstlisting}[language=Python,numbers=left,basicstyle={\ttfamily},breaklines=true,tabsize=6]
class GraphConvolution(Module):     

    def __init__(self, in_features, out_features, bias=True):
        super(GraphConvolution, self).__init__()
        self.in_features = in_features
        self.out_features = out_features
        self.weight = Parameter(torch.FloatTensor(in_features, out_features))
        if bias:
            self.bias = Parameter(torch.FloatTensor(out_features))
        else:
            self.register_parameter('bias', None)
        self.reset_parameters()

    def reset_parameters(self):
        stdv = 1. / math.sqrt(self.weight.size(1))
        self.weight.data.uniform_(-stdv, stdv)
        if self.bias is not None:
            self.bias.data.uniform_(-stdv, stdv)

    def forward(self, input, adj):
        support = torch.mm(input, self.weight)
        output = torch.mm(adj, support)
        if self.bias is not None:
            return output + self.bias
        else:
            return output
    def __repr__(self):
        return self.__class__.__name__ + ' (' \
            + str(self.in_features) + ' -> ' \
            + str(self.out_features) + ')'
\end{lstlisting}

\end{algorithm}
\begin{algorithm}
\caption{\label{alg:Graph-Convolutional-Autoencoder}Graph Convolutional Autoencoder}

\begin{lstlisting}[language=Python,numbers=left,basicstyle={\ttfamily},breaklines=true,tabsize=2]
class GCAE(nn.Module):
    def __init__(self, nfeat, nhid, nlatent, depth=10):
        super(GCAE, self).__init__()
        self.nfeat = nfeat
        self.nhid = nhid
        self.nlatent = nlatent
        self.encoder_gc_init = GraphConvolution(nfeat, nhid, bias=False)
        self.encoder_gc = [GraphConvolution(nhid, nhid, bias=False) for _ in range(depth)]
        self.encoder_gc_final = GraphConvolution(nhid, nlatent, bias=False)
        self.decoder_gc_init = GraphConvolution(nlatent, nhid, bias=False)
        self.decoder_gc = [GraphConvolution(nhid, nhid, bias=False) for _ in range(depth)]
        self.decoder_gc_final = GraphConvolution(nhid, nfeat, bias=False)

    def encode(self, x, adj):
        x = F.relu(self.encoder_gc_init(x, adj))
        for gc in self.encoder_gc:
            x = F.relu(gc(x, adj))
        z = torch.sigmoid(self.encoder_gc_final(x, adj))
        return z

    def decode(self, z, adj):
        x = F.relu(self.decoder_gc_init(z, adj))
        for gc in self.decoder_gc:
            x = F.relu(gc(x, adj))
        x = F.relu(self.decoder_gc_final(x, adj))
        return x

    def forward(self, x, adj):
        z = self.encode(x, adj)
        recon = self.decode(z, adj)
        return recon 
\end{lstlisting}

\end{algorithm}

 The model expects input data $x$ and $adj$, each being torch tensors,
with shapes of $n\times v$ and $n\times n$ respectively where $n$is
the length of the given code sequence, and $v$ is the size of the
vocabulary. This represents the sequence of one-hot representations
of the code, and the adjacency matrix for the graph. Then depending
on the $depth$ parameter, this input is passed through a number of
GraphConvolution layers, where the hidden states have shape $n\times h$,
where $h$ is the hidden size parameter, this represents the hidden
state value for each node. After passing through the encoding layers,
the latent value is generated by passing through a sigmoid activation
and producing a latent variable sequence of size $n\times l$ representing
the code sequence. Note through all of these Graph Convolution layers,
the same adjacency matrix is used unchanged, which is the same strategy
used in the original work for deep Graph Convolutional Networks. Then
a symmetric process is used in decoding the latent variable to a reconstructed
code sequence tensor. As can be seen clearly, the model actually is
independent of the size of the sequence, relying on the graph properties
to propagate information. This is a benefit meaning there does not
need to be any padding and the same network can be trained directly
on the various sized methods in the dataset. However because of this
that means the latent space is also dependently sized, as opposed
to a more standard fixed vector. This will mean that the actual latent
dimension for this experiment to be meaningful will have to be very
small, in order to sufficiently compress the original sequence and
make the latent representation meaningful and useful.

For the experiment, models were trained with the same procedure and
with the same hyperparameters, varying only the adjacency matrix.
A hidden size of 32, zero additional depth layers, and a latent size
of 4 were chosen, which while small by many deep learning standards
were in line with he size of the dataset and compute resources available,
and helped reduce model size and variance. In particular the latent
size of 4 allows for visualization of program encodings, which is
among the goals of the work. The training procedure utilized a Cross
Entropy Loss objective function, which is common in autoencoders and
in this case also frames the autoencoder as a form of classification,
classifying the token value for each position. The models were also
regularized using $l_{2}$ regularization with a value of $10^{-5}$,
and trained using Adam optimization\cite{2014arXiv1412.6980K}
with a learning rate of $10^{-3}$. Each of the three scenarios of
adjacency matrix were trained over the dataset of $13000$ methods
over $5$ epochs. These are very small models by deep learning standards.
This is partially due to time and computational restrictions but also
due to the heavily abstracted and simplified nature of the task. And
so while it may be possible to achieve more interesting results in
future works using larger networks on a less abstracted problem, this
simplified problem is still able to show a few key results in the
nature of how graph convolutional networks are able to encode programs
and exposes certain limitations even given the small scale implementation.

\chapter{Evaluation and Analysis}

\begin{figure}
\noindent \begin{centering}
\caption{\label{fig:Training-Curves}Training Curves}
\par\end{centering}
\noindent \centering{}\hspace{-2cm}\includegraphics[scale=0.85]{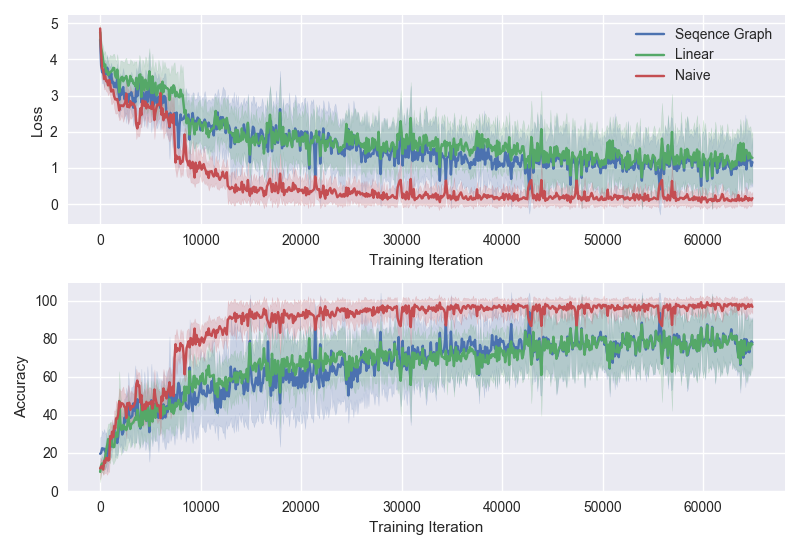}
\end{figure}
\begin{table}
\noindent \begin{centering}
\caption{\label{tab:Test-Metrics}Test Metrics and Standard Deviations}
\par\end{centering}
\vspace{0.5cm}
\noindent \centering{}%
\begin{tabular}{ccccc}
\toprule 
 & Mean Loss & $\sigma$ & Mean Accuracy & $\sigma$\tabularnewline
\midrule
\midrule 
Sequence Graph & 0.87364 & 0.743 & 83.0\% & 15.0\%\tabularnewline
\midrule 
Linear Graph & 0.92246 & 0.788 & 83.5\% & 14.8\%\tabularnewline
\midrule 
Naive Graph & 0.07852 & 0.146 & 98.5\% & 2.9\%\tabularnewline
\bottomrule
\end{tabular}
\end{table}

After running the training procedure described above, Figure \ref{fig:Training-Curves}
shows the training curves for all three models showing both the training
loss and accuracy (as measured as percent of tokens correctly reconstructed).
Table \ref{tab:Test-Metrics} shows the same metrics on unseen test
data. 
\begin{table}
\caption{\label{tab:Example-Reconstructions}Example Reconstructions}

\begin{adjustbox}{center}%
\begin{tabular}{ll}
\toprule 
Original Code & 
\begin{lstlisting}[basicstyle={\tiny\ttfamily}]
method ( int n ) { int id = 1 ; for ( int i = 2 ; i <= n ; i ++ ) { id *= i ; } return id ; }
\end{lstlisting}
\tabularnewline
\midrule 
Sequence Graph & 
\begin{lstlisting}[basicstyle={\tiny\ttfamily}]
method ( id id ) { ; id = ; } } id id , id return return id id id ; id id ) { ) ( { ; } return id ; }
\end{lstlisting}
\tabularnewline
\midrule 
Linear Graph & 
\begin{lstlisting}[basicstyle={\tiny\ttfamily}]
method ( id id ) { ; ) ) ) ; id ( id id id ) ) ) ) ) ; id ) id id id id id ; } return id ; }
\end{lstlisting}
\tabularnewline
\midrule 
Naive Graph & 
\begin{lstlisting}[basicstyle={\tiny\ttfamily}]
method ( int :: ) { int id = :: ; for ( int i = :: ; i :: :: ; i ++ ) { id :: i ; } return id ; }
\end{lstlisting}
\tabularnewline
\bottomrule
\end{tabular}\end{adjustbox}
\end{table}
 Table \ref{tab:Example-Reconstructions} shows the resultant reconstructed
text of the different models for an unseen test method. As these results
show, performance is substantially worse than expected, and in fact
shows a trend precisely opposite to what is expected. The hypothesis
predicted that the sequence graph result would outperform linear which
would outperform naive graphs. In fact the naive graph vastly outperformed
the other models, and there was little difference between the sequence
and linear graph models. Therefore this work is forced to reject the
initial hypothesis and come to some explanation of the results. And
in the spirit of science, a negative result is as good as a positive
result. The rest of this section will aim to analyze these results,
and understand what exactly happened.

One key component that becomes clear upon analysis is the essential
similarity between the linear and sequence graph models. When looking
closely at most of the natural code in the dataset it can be seen
that the vast majority of methods are actually just simple imperative
executions of small chunks of code. This makes sense within the object
oriented paradigm of Java where each method aims to do one very limited
function. What this means then is that this specific kind of static
analysis does not actually induce a sufficiently large difference
for the vast majority of programs such that these networks can learn
the difference within a reasonable amount of time over the size of
the data collected. In short, one of the takeaways is that \emph{the
vast majority of natural methods do not have very complicated control
flows }and therefore models relying on information from control flow
graphs for small segments of object oriented code (which is the vast
majority of how real code is structured) have strong limitations.

Furthermore looking at example reconstructions such as in table \ref{tab:Example-Reconstructions},
another pattern becomes clear. The kinds of predictions the poorly
performing models are making are mostly in 'hedging their bets' and
guessing many common terms in a row to get at least some tokens correct.
Part of this can be attributed to the way this test was conducted,
parsing more finely into individual tokens more heavily weights much
of the minutiae of the program, and therefore these models learned
statistically useful but practically meaningless patterns. This kind
of situation may occur when the actual noise of the input signal is
too great, and so the models learn to be more biased. This is a likely
explanation, due to the mechanics of how graph convolution works.
The transformations in Graph Convolutional Networks learn to propagate
information fully locally between connected nodes (including the augmented
self loops). In order for this to work it essentially aggregates all
local connections together within the same weight. What this means
is that in the naive case where only self loops exist the network
is able to learn how each token can be encoded purely based on itself,
building what amounts to a word embedding. However when introducing
any linear connections, the network cannot tell the difference between
an edge connecting the previous token and the self loop, or even further
if there is an additional edge from the control flow, information
propagates in the exact same way. What this amounts to is anything
other than the self loops just introducing noise into the system,
making the network more biased. Furthermore the purely local nature
of these transformations means that the network cannot learn the global
high level structural information that is essentially the only information
that is maintained through the abstraction preprocessing. This would
be fixed in more traditional convolution by adding depth to the network,
however because of the aggregation property this essentially just
multiplies the noise effect when attempted. In addition experiments
where the identity augmentation is removed from the graph convolutional
models also produce substantially worse performance due to the inability
to propagate any information in the same node from layer to layer
making the model nothing more than trying to predict a token two or
three tokens in the future. Essentially graph convolutional networks
perform best under a number of conditions, namely strong local correlation,
node agnostically, and much high density of connections than are found
in control flow graphs. Due to this experiment lacking those properties,
graph convolutions are a poor fit for representing programs in this
way. The naive graph model is able to avoid these issues by having
only the identity connections, and therefore being able to actually
learn a consistent way to encode. However this encoding is not actually
that insightful, as it more or less acts as a pure encoding of each
individual token, containing no information about the program globally.
And so while it performs well it is not very interesting. Essentially
the main result of this analysis is that \emph{graph convolutional
networks do not encode information from control flow graphs well}
and that in the world of program synthesis too much is lost when trying
to abstract away program specifics to fit within a purely neural network
based model.

\chapter{Conclusions and Future Work}

The primary result of this study is that using the popular notion
of Graph Convolutional Networks, augmentation with control flow graphs
have no discernible gain in program embedding performance, and in
fact graph convolutions in general are a poor choice for the task
currently when compared to other results using linguistic models.
This has implications in the development of geometric deep learning,
where it would imply that while existing local methods are useful
in certain applications where the graphs in question imply more basic
correlation between connected nodes, there is a lack of methods interpreting
nuanced global properties of the graphs. In the field of program synthesis
it also implies that more purely linguistic solutions that can take
advantage of more sophisticated models like attention are preferable
to augmentation of less powerful models with formal analysis. And
even further that the necessary abstractions required in order to
fit programs into a deep learning framework remove too much nuance
and thus purely statistical methods are not likely to be effective
in program synthesis. Therefore it also further supports the use of
CEGIS style frameworks for program synthesis, where the oracle consisting
of statistical or deep learning models and the formal verifier are
more separated and each task is more individually optimized, as the
relevant information toward those tasks do not substantially overlap.
Finally the most valuable conclusions from such a strong negative
result, given that this is the first empirical evaluation of its kind
and the many computational limitations and strong abstractions, is
the need for replication. However if these results are replicated,
it would certainly have an effect on the direction of future program
synthesis research.

\bibliographystyle{apalike}
\bibliography{refs} 
\end{document}